\documentclass{article}
	
	\usepackage{amsmath}
	\usepackage{amsfonts}
	\usepackage{amsthm}
	\usepackage{mathtools}
	\usepackage{setspace}
	\usepackage{fullpage}
	
	\usepackage{courier}
	
	\usepackage{algpseudocode}
	
	\usepackage{tikz}
	\tikzset{node distance=2cm, auto}
	
	\newtheorem{mydef}{Definition}
	\newtheorem{mythm}{Theorem}

	\numberwithin{equation}{section}
	\numberwithin{mydef}{section}
	\numberwithin{mythm}{section}
	
	\usetikzlibrary{shapes,shadows,arrows}
	
	\tikzstyle{startstop} = [ellipse, minimum width=1cm, minimum height=1cm, text centered, draw=black]
	\tikzstyle{io} = [trapezium, trapezium left angle = 70, trapezium right angle = 110, minimum width=1cm, minimum height=1cm, text centered, draw=black]
	\tikzstyle{process} = [rectangle, minimum width=1cm, minimum height=1cm, text centered, draw=black]
	\tikzstyle{decision} = [diamond, minimum width=1cm, minimum height=1cm, text centered, draw=black]
	\tikzstyle{arrow} = [thick,->,>=stealth]
	
	\title{ \textbf{A similarity criterion for sequential programs using truth-preserving partial functions}}
		
	\author{Abhinav Aggarwal\\
				\textit{Department of Computer Science and Engineering}\\
				Indian Institute of Technology, Roorkee}
				
	\date{}
	
\begin{document}
	
	\maketitle
		
	\begin{abstract}
		
	The execution of sequential programs allows them to be represented using mathematical functions formed by the composition of statements following one after the other. Each such statement is in itself a partial function, which allows only inputs satisfying a particular Boolean condition to carry forward the execution and hence, the composition of such functions (as a result of sequential execution of the statements) strengthens the valid set of input state variables for the program to complete its execution and halt succesfully. With this thought in mind, this paper tries to study a particular class of partial functions, which tend to preserve the truth of two given Boolean conditions whenever the state variables satisfying one are mapped through such functions into a domain of state variables satisfying the other. The existence of such maps allows us to study isomorphism between different programs, based not only on their structural characteristics (e.g. the kind of programming constructs used and the overall input-output transformation), but also the nature of computation performed on seemingly different inputs. Consequently, we can now relate programs which perform a given type of computation, like a loop counting down indefinitely, without caring about the input sets they work on individually or the set of statements each program contains.
		
	\end{abstract}
	
	\section{Introduction}
	
	Computer programs can be broadly classified into being sequential or parallel, based on the nature of computation of various modules in it. If the statements in a program are executed one after the other in the order they are written in the code (allowing control flow structures like loops and conditional jumps), we say that a program executes sequentially. If, however, the program executes different modules on different machines (running independently of each other and in parallel) and then combines the results obtained, we say that it is a parallel program. Our analysis is limited to the programs of the first kind, i.e. the ones for which the execution of statements depend on the order of their appearance in the code. This allows for using the concept of composition while dealing with multiple statements. If we represent a statement $S$ by an equivalent function (essentially a transformation map), say $S(\alpha)$, where $\alpha$ is the set of state variables on which $S$ operates, then the execution of statements $S_{1}$ and $S_{2}$ (in this order) can be written as the composition $S_{2} \left( S_{1}(\alpha) \right)$. The idea behind this composition is the feeding of inputs to $S_1$ and then forwarding the outputs, thus obtained, to the input of $S_2$. The final outcome of this composite transformation is then referred to as the result of the execution of statements $S_1$ and $S_2$, in this order. Although many approaches to formally represent computational steps of an algorithm exist, the partial function model becomes particularly useful to study mathematical properties of the underlying code. In the frequent case of conditional execution of a given statement $S$ based on some Boolean condition $C$, (as in the case of \textit{if-conditionals}), the use of partial function simplifies the study of denotational semantics through a representation using an equivalent partial function of the form
	
	\begin{equation}
	S'(x)=
	\begin{cases}
	S(x) & \text{if }C \text{ is true}\\
	\text{\emph{undefined}} & \text{otherwise}
	\end{cases}
	\end{equation}  
	
	Similarly, the \textit{while-loops} and other constructs in a typical sequential program can be represented using their equivalent mathematical functions (total or partial) and their appropriate compositions. 
	
	\paragraph*{Using partial functions}
	\label{chap:UsingPartialFn}
	Let $A$ and $B$ be two sets and $f:A \to B$ be a partial bijection. Then the definition of $f$ on only a selected few elements of $A$ can be interpreted as $f(x)$ being defined on $x$ only if $x$ satisfies some condition $C$, and remains undefined otherwise. We represent the set of all $x \in A$ that satisfy the condition $C$ by $[A]_{C}$. More precisely, the set $[A]_{C} \subseteq A$ is a restriction of $A$ with respect to the condition $C$. Thus, $Dom(f)=[A]_{C}$ and $Im(f)=f(Dom(f)) \subseteq B$. The application of $f$ to this subset of $A$ is denoted by $f([A]_{C})$, meaning that $f$ is only defined for those elements in $A$ for which $C$ is true. Note that if $C$ is \textit{False} for all elements in $A$, then the set $[A]_{C}=\Phi$, the empty set. Also, for each subset $S$ of $A$, we can find a condition $C$ such that $S = [A]_{C}$. The maximum number of subsets of $A$ is $2^{|A|}$, which means that the maximum number of Boolean conditions which can produce distinct restrcitions of $A$ can be $2^{|A|}$. Thus, the set $\mathcal{C}_A$ of all possible Boolean conditions that can ne defined on the elements of $A$ is partitioned into $2^{|A|}$ equivalence classes, where the conditions belonging to the same class produce the same restriction of the set $A$, whereas conditions across the classes produce different ones. \\
	
	For any set $A$, we can define a conditional identity function, $id_{C}:A \to [A]_{C}$ as 
	
	\begin{equation}
	\label{Eq:idc}
	id_{C}(x)=
	\begin{cases}
	x & \text{if } x \in [A]_{C}\\
	\text{\emph{undefined}} & \text{otherwise}
	\end{cases}
	\end{equation}
	
	We can extend this concept to functions, where the application of a given function $f:A \to B$ is conditioned on a Boolean expression $C \in \mathcal{C}_A$. We denote this restricted function application by $f_{C}$ and define it as:
	
	\begin{equation}
	\label{Eq:partFunc}
	f_{C}(x)=
	\begin{cases}
	f(x) & \text{if } x \in [A]_{C}\\
	\text{\emph{undefined}} & \text{otherwise}
	\end{cases}
	\end{equation}
	
	Thus, we have $f_{C} \simeq f([A]_{C})$. The two notations can be used interchangably. \\
	
	Now, consider the composition of $f_{C_{1}}:A \to B$ with another function $g_{C_{2}}:B \to C$ to get $h_{C_{3}}:A \to C = (g \circ f)$. The aim is to find $C_{3}$ in terms of $C_{1}$ and $C_{2}$. We can write $h_{C_{3}}$ as
	
	\begin{equation}
	h_{C_{3}}(x)=
	\begin{cases}
	(g \circ f_{C_{1}})(x) & \text{if } f_{C_{1}}(x) \in [B]_{C_{2}}\\
	\text{\emph{undefined}} & \text{otherwise}
	\end{cases}
	\end{equation}
	
	This is quivalent to writing
	
	\begin{equation} \label{eq:hc3}
	h_{C_{3}}(x)=
	\begin{cases}
	(g \circ f)(x) & \text{if } x \in [A]_{C_{1}} \text{ and } f(x) \in [B]_{C_{2}}\\
	\text{\emph{undefined}} & \text{otherwise}
	\end{cases}
	\end{equation}
	
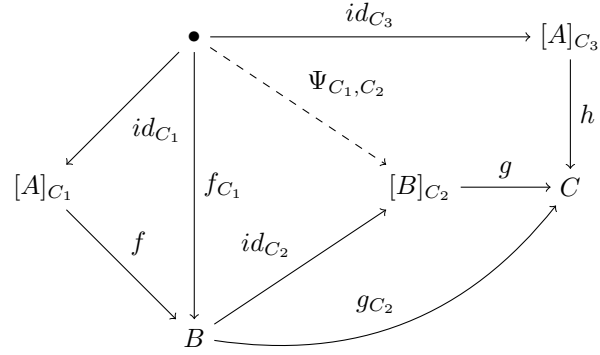
\begin{figure}[htp]
	\centering
	\begin{tikzpicture}
	\node (A) {$\bullet$};
	\node (Ac) [below of=A, left of=A] {$[A]_{C_{1}}$};
	\node (B) [below of=Ac, right of=Ac] {$B$};
	\node (Bc) [node distance=5cm, right of=Ac] {$[B]_{C_{2}}$};
	\node (Ac3) [above of=Bc, right of=Bc] {$[A]_{C_{3}}$};
	\node (C) [right of=Bc] {$C$};
	
	\draw[->] (A) to node {$id_{C_{1}}$} (Ac);
	\draw[->] (Ac) to node {$f$} (B);
	\draw[->] (B) to node {$id_{C_{2}}$} (Bc);
	\draw[->] (Bc) to node {$g$} (C);
	\draw[->] (Ac3) to node {$h$} (C);
	\draw[->] (A) to node {$id_{C_{3}}$} (Ac3);
	\draw[->] (A) to node {$f_{C_{1}}$} (B); 
	\draw[->, bend right] (B) to node {$g_{C_{2}}$} (C);
	\draw[->, dashed] (A) to node {$\Psi_{C_{1},C_{2}}$} (Bc);  
	\end{tikzpicture}
	\caption{Commutative diagram for composition of partial functions} 
	\label{fig:Fig1}
	\end{figure}

	Thus, the condition $C_{3}$ is equivalent to ($x \in [A]_{C_{1}} \text{ and } f(x) \in [B]_{C_{2}}$). Here, whenever $C_{1}$ is true for a given $x$, we must have $C_{2}$ true for the corresponding $f(x)$. This fact has been demostrated using Fig. \ref{fig:Fig1}.\\
	
	The diagram commutes whenever \eqref{eq:hc3} holds. Notice the dashed arrow for a function $\Psi_{C_{1},C_{2}}:A \to [B]_{C_{2}}$, given as $\Psi_{C_{1},C_{2}} = (id_{C_{2}} \circ f_{C_{1}})$. Thus, $\Psi_{C_{1},C_{2}}$ is a partial function from $A$ to $B$, defined as: 
	
	\begin{equation}
	\Psi_{C_{1},C_{2}}(x)=
	\begin{cases}
	f(x) & \text{if } x \in [A]_{C_{1}} \text{ and } f(x) \in [B]_{C_{2}}\\
	\text{\emph{undefined}} & \text{otherwise}
	\end{cases}
	\end{equation}
	
	Intuitively, $\Psi_{C_1,C_2}(x)$ preserves the satisfiability of condition $C_1$ across the mapping, $f$, through the satisfaction of $C_2$. In a way, given that some $y \in B$ satisfies $C_2$ and an inverse image of $y$ exists under $f$, this pre-image $f^{-1}(y)$ is guaranteed to satisfy $C_1$. This property of the so-called \textit{truth preservation} of $C_1$ by $C_2$ is important to study the nature of inputs two seemingly different programs operate on as well as the kind of transformation they simulate. The next section discusses these functions in greater detail.
	
	\subsection{Truth Preserving Functions}
	
	The function, $\Psi_{C_{1},C_{2}}$, essentially takes elements from $A$ into the set $B$, preconditioned on $C_{1}$ and postconditioned on $C_{2}$. By this, we mean that the mapping from $A$ to $B$ is only done for elements satisfying $C_{1}$ (the check being performed prior to the transformation) and once the required mapping to the set $B$ has been completed, all elements not satisfying $C_{2}$ are filtered out. Essentially, for each element $x \in A$ that satisfies $C_{1}$, we finally have $\Psi_{C_{1},C_{2}}(x)$ satisfying $C_{2}$, and for each element $x' \in A$ that does not satisfy $C_{1}$, the function $\Psi_{C_{1},C_{2}}(x')$ remains undefined. In a way, through the three partial functions that compose together to form $\Psi_{C_{1},C_{2}}(x')$, the truth of $C_{1}$ seems to be preserved in $C_{2}$. However, the map does not alter the elements of $B$ which satisfy $C_{2}$ in any way, and thus $C_{2}$ becomes a weaker condition than $C_{1}$ here. There can exist elements in $B$ that satisfy $C_2$ but do not have a pre-image under $f$. We can then demostrate this fact using the logical deduction $C_{1} \vdash C_{2}$, since the falsehood of $C_{1}$ renders the truth value of $C_{2}$ unimportant. In this context, we can refer to $\Psi_{C_{1},C_{2}}$ as a \textit{truth preserving partial function} from $C_{1}$ to $C_{2}$, with respect to the sets $A$ and $B$. This is formalized in the definition below.\\
	
	\begin{mydef}
	\label{def:truth_Preserv}
	Let $A$ and $B$ be two sets and $C_{A} \in \mathcal{C}_A$ and $C_{B} \in \mathcal{C}_B$ be two Boolean conditions that are defined on the elements of $A$ and $B$ respectively. Let $f:A \to B$ be a partial injective (one-one) function, mapping only those elements of $A$ that satisfy $C_{A}$. Then the partial function $\Psi_{C_{A},C_{B}}:A \to B \simeq (id_{C_{B}} \circ (f \circ id_{C_{A}}))$ is called a \textbf{Truth Preserving Partial Function} from $A$ to $B$ with respect to $C_A$ and $C_B$.\\
	\end{mydef}
	
	In this context, $C_{A}$ seems to entail $C_{B}$, i.e. $C_{A} \vdash C_{B}$, and hence, $C_{A}$ logically implies $C_{B}$. We use the notation $C(x)$ to represent the truth value obtained on checking the condition $C \in \mathcal{C}_A$ against the assignments of its variable using a value $x$ in the set $A$. Another way to represent this implication is by using bounded quantifiers:
	\begin{equation}
	\left[ \forall x \in Dom(id_{C_{A}}:A \to A) \right] C_{A} \quad \vdash \quad \left[ \forall y \in Im(id_{C_{B}}:Im(f) \to B) \right] C_{B}
	\end{equation}
	
	A partial function $f:A \to A$, for a given set $A$ and a Boolean expression $C_{A} \in \mathcal{C}_A$ is truth preserving if, according to Def \ref{def:truth_Preserv}, it can be written as $f=(id_{C_{A}} \circ (g \circ id_{C_{A}}))$ for some function $g:[A]_{C_{A}} \to A$. In other words, whenever $C_{A}$ is true for some $x \in A$, it is also true for $f(x)$. This means that $C_{A}$ must be true for all elements in the orbit of $x$ under $f$, i.e. $\{x,f(x),f(f(x)),\dots\}$. Consider the code snippet as given in Fig. \ref{Fig:loop}. Assume that the loop is entry controlled and has a single exit point. Thus, no possibility of exit from the loop while $C_A$ still being true is possible.\\
	
	\begin{figure}[htp]
	\texttt{
	\begin{algorithmic}
	\State Initialize $x \in A$
	\While{$C_{A}$ is \emph{True} on $x$}
		\State Set $x \gets f(x)$
	\EndWhile
	\State Print $x$
	\end{algorithmic}
	}
	\caption{A general \texttt{while}-loop} \label{Fig:loop}
	\end{figure}
	
	The output is seen only for those values of $x \in A$ for which the condition $C_{A}$ is false before entry in the loop. Once the loop is entered, there is no coming out because $C_{A}$ will then be true for all values $x$ will take. Thus, we enter into a non-terminating computation for all $x \in [A]_{C_A}$. The question that needs to be answered here is \emph{How to determine if a given function $f$ is truth preserving for a given Boolean condition $C \in \mathcal{C}_A$?} \\
	
	We can visualize truth preserving functions as mapping their arguments into a set, the elements of which belong to $Dom(f)$, i.e. only when $f$ maps $x$ into a set such that $f$ is also defined on the value $f(x)$, can we apply $f$ again and continue with our execution. This has to happen for every input argument $f$ takes and hence, we must have the range of $f$ to be a subset of its domain, i.e. $Im(f) \subseteq Dom(f)$. In such a case, the repeated application of $f$ will always be defined on all valid inputs to $f$ and the truth of the Boolean condition responsible for the restriction of $Dom(f)$ to $Im(f)$ will then be preserved by $f$ indefinitely. For the code snippet above, this condition can be written as $Im(f_{C_A}) \subseteq Dom(f_{C_A})$. To check this condition, it suffices to determine if $Im(f) \subseteq [A]_{C_A}$, for if this were true, then any argument to $f$ will always produce $f(x) \in [A]_{C_A}$ as required. \\
	
	For cases where $Im(f) \not\subseteq [A]_{C_A}$, that is $f$ is not a truth preserving function $\forall x \in [A]_{C_A}$, the elements in the orbit of $f$, i.e. $\{ x, f(x), f(f(x)), \dots \}$ may not all satisfy  $C_A$. Rather, for every $x \in [A]_{C_A}$, there may only be a finite number of elements in this orbit which will satisfy $C_A$. This provides a bound on the number of times the above loop would execute. The upcoming subsections analyze such cases in greater detail. 
	
	\subsection{Order of truth preservation}
	
	To study the behavior of functions with respect to a given Boolean condition, we define the notion of \textit{order} of the \textit{truth preservation} of a given function as follows: (Assume that $\mathbb{N} = \{0,1,2,3,4,\dots\}$)\\
	
	\begin{mydef}
	\label{def:order_truth_preserv}
	Let $A$ be a set and $C_{A} \in \mathcal{C}_A$ be a Boolean condition. Let $f:A \to A$ be a partial function such that $Dom(f)=[A]_{C_{A}}$ and $\exists n_{x} \in \mathbb{N}$ such that $n_{x}$ is the smallest number for which $C_{A} \left( f^{(n_{x}+1)} (x) \right)$ first becomes false. Let $m \in \mathbb{N}$ be the minimum of these $n_{x}$'s for all $x \in Dom(f)$, i.e. 
	\begin{equation}
	m=\min_{x \in [A]_{C_{A}}}  {n_{x}}
	\end{equation}
	In this case, call $n_{x}$ the \textbf{order of $\textbf{x}$} with respect to the truth preservation by the function $f$ of the condition $C_{A}$, denoted by $\Theta_{f,C_{A}}(x)=n_{x}$, and call $m$ the \textbf{order of $\textbf{f}$} for the truth preservation of $C_{A}$ with respect to the set $A$, denoted by $\Theta_{C_{A}}(f)=m$.\\
	\end{mydef} 
	
	To understand this definition, let us consider the same code snippet in Fig. \ref{Fig:loop}. Note that $Dom(f)=[A]_{C_A}$ and the loop condition is $C_A$. This simplifies our analysis because we can now be sure that the execution of $f$ will only be done for elements in its domain. We will shortly discuss the case where $C_A(x)$ will be true but $f(x)$ will not be defined.\\
	
	The statement of Def. \ref{def:order_truth_preserv} asserts that the loop construct, as shown previously, must execute atleast $(\Theta_{C_{A}}(f)+1)$ times because the condition $C_{A}$ will become false for the first time after atleast these many iterations of the loop. Thus, $m+1$ provides a definite lower bound on the loop execution count. For a given element $x$, chosen during the initialization step, the loop condition becomes false after exactly $n_x +1$ iterations and the program terminates after printing $f^{(n_x +1)}(x)$ as the output. This condition can easily be checked by verifying that if for a given $x \in [A]_{C_A}$, we have $f^{(n)}(x) \in [A]_{C_A}$ for all $0 \leq n \leq m $ as well, then the value of $n_x$ is $m$ at least. \\
	
	An interesting phenomenon that is encountered frequently in this scenario is that of $k$-periodicity. A set of points $\{x_0, x_1, x_2, \dots, x_{k-1} \}$ is said to be $k$-periodic with respect to a function $f$ if 
	
	\begin{equation*}
	f(x_i) = x_{(i+1) \bmod k}, \quad 0 \leq i < k 
	\end{equation*}  
	
	The orbit of $x_i$ is thus given as
	
	\begin{equation*}
	\emph{orbit}(x_i) = \{ x_i, x_{(i+1) \bmod k}, x_{(i+2) \bmod k}, \dots, x_{(i+k-1) \bmod k}, x_i, \dots \}
	\end{equation*}
	
	We have $x_i = f^{(k)}(x_i)$ true for all $x_i$ in the set of $k$-periodic points. For the truth preservation order, however, not all $k$-points may satisfy the given condition $C_A$. In case they all do, then it is trivial to see that $n_x = \infty$ for all of them. In all other cases, the order is limited above by $k$. The special case here is of fixed points, for which $k=1$, i.e. points which satisfy $f(x)=x$. These points either do not satisfy $C_A$ at all, or have an infinite order. Another interesting case is of functions of the form $g = (id_{C_A} \circ f)$ for some given $f$. Clearly, the type of these functions is $(A \to [A]_{C_A})$. If $f$ is total, then for all $x \in A$, we will have $C_A(g^m(x))$ to be true for all $m \geq 0$ and hence, the order of $g$ becomes infinity. \\ 
	
	An upper bound on the number of loop iterations is given by 
	\begin{equation}
	l=\max_{x \in [A]_{C_{A}}} {n_{x}} 
	\end{equation}
	This value can be referred to as the \textbf{truth preservation limit} of $f$ for the condition $C_{A}$ with resepct to the set $A$, denoted by $\hat{\Theta}_{C_{A}}(f)=l$. Since this is an upper limit on the number of times application of $f$ to $x$ will preserve the truth of $C_{A}$, the number of loop iterations (as in the code snippet above) will never exceed $l+1$. The following example illustrates the calculation of order and limit for a given function and a given Boolean condition:\\
	
	\texttt{
	\begin{algorithmic}
	\State Initialize $x \in \{1,2,3,4,\dots,15\}$
	\While{x < 10}
		\State Set $x \gets x+2$
	\EndWhile
	\State Print $x$ \\
	\end{algorithmic}
	}
	
	For this example, we have $A = \{ 1,2,3,\dots,15 \}$, $C_A = [x < 10]$, $[A]_{C_A} = \{ 1,2,3,\dots,9 \}$ and $f(x)=x+2$. The orders of various elements in $[A]_{C_A}$ is given in the table below:
	
	\begin{center}
	\begin{tabular}{| l | *{9}{c |}}
	\hline
	$x$ & 1 & 2 & 3 & 4 & 5 & 6 & 7 & 8 & 9 \\ \hline
	$n_x$ & 4 & 3 & 3 & 2 & 2 & 1 & 1 & 0 & 0\\ \hline
	\end{tabular}
	\end{center}
	Thus, we have $m = 0$ and $l = 4$, i.e. the loop runs for at most $5$ times, when the value of $x$ chosen in the initialization step is $1$, and at least once when $x$ is chosen to be either $8$ or $9$.\\
	
	We can generalize the definition of truth preservation order and limit to accomodate for all $x \in A$, instead of restricting ourselves to only $x \in [A]_{C_A}$. This will make further analysis easier by not limiting the results to restricted domains. Note that in case $x \not\in [A]_{C_A}$, the definition of $n_x$ requires $C_A(f^{(n_x+1)}(x))$ to be \emph{False} for the first time. This means that we must have $n_x = -1$ for all $x \in A \backslash [A]_{C_A}$. Hence, the new definition of truth preservation order can be given as
	\begin{equation}
	m = \max \left( -1, \min_{x \in [A]_{C_A}} n_x \right)
	\end{equation}
	
	With this new definition, a value of $m = -1$ will assert that for no $x \in [A]$ do we have $C_A(f(x)) = \emph{True}$, which the earlier definition of $m$ would not have been able to conclude. Similar to this modified definition of truth preservation order, the truth preservation limit can now be defined as
	\begin{equation}
	l = \max \left( -1, \max_{x \in [A]_{C_A}} n_x \right) = \max_{x \in A} n_x
	\end{equation} 
	
	The order and limit for any function $f$ are constants, theoretically. However, in a practical scenario, working on a $t$-bit computer can change the behavior completely. The bit precision that comes into picture restricts the representability of any real number to numbers modulo $2^t$, and hence, the order and limit for $f$ may change. Many other factors can affect this truth preservation under changing circumstances. The next subsection enumerates some important ones.
	
	\subsection{Factors affecting truth preservation order and limit}
	
	For any function $f$, given $C_{A}$, the truth preservation order $m$ and the limit $l$, we have a lower and upper bound, respectively, on the number of iterations of the loop in the code snippet as in Fig. \ref{Fig:loop}. Thus, if it takes $T(x)$ number of loop iterations for a given $x \in A$, then we have 
	
	\begin{subequations}
	\begin{align}
	T(x)=0 \quad \text{if } C_{A}(x) \text{ is \textit{False}}\\
	m+1 \leq \left( T(x)=n_{x} \right) \leq l+1 \quad \text{if } C_{A}(x) \text{ is \textit{True}}
	\end{align}
	\end{subequations} 
	
	Let us now see how these bounds can change under various situations. 
	\subsubsection{Condition Strengthening} If the condition $C_{A}$ is strengthened to $C_{A}'$ such that $C_{A}' \vdash C_{A}$, then we have a case when the looping criterion has been made stricter and the number of iterations is likely to decrease. This is because of a possible reduction in the number of $x \in A$ that will satisfy $C_A'$. However, when $C_{A}$ first becomes \textit{False}, we must have $C_{A}'$ to be \textit{False} as well, but not the viceversa. When $C_{A}'$ first becomes false, $C_{A}$ may still be true and hence, the new truth preservation order, $m'$ is bounded above by $m$. 
	\begin{equation}
	\Theta_{C_{A}'}(f) \leq \Theta_{C_{A}}(f) \quad \text{if } C_{A}' \vdash C_{A}
	\end{equation}
	For the case of limits, we see that the maximum number of loop iterations also reduces since the falsehood of $C_{A}'$ does not gaurantee the same for $C_{A}$. Hence, we have:
	\begin{equation}
	\hat{\Theta}_{C_{A}'}(f) \leq \hat{\Theta}_{C_{A}}(f) \quad \text{if } C_{A}' \vdash C_{A}
	\end{equation}   
	Thus, the number of loop iterations is likely to reduce on condition strengthening, as is expected. However, there may be situations where the matter of concern is not the reduction per se but the amount of reduction in these bounds. For this, define the following two quantities:
	
	\begin{equation}
	\label{eq:delta_order}
	\underbrace{\Delta \Theta_{C_{A},C_{A}'} (f)}_\text{Order relaxation} = \Theta_{C_{A}}(f) - \Theta_{C_{A}'}(f)
	\end{equation}
	
	\begin{equation}
	\label{eq:delta_limit}
	\underbrace{\Delta \hat{\Theta}_{C_{A},C_{A}'} (f)}_\text{Limit relaxation} = \hat{\Theta}_{C_{A}}(f) - \hat{\Theta}_{C_{A}'}(f)
	\end{equation}
	
	Clearly, higher values of both relaxations provide smaller bounds on the number of loop iterations. However, we still have no information about the change in $n_x$ for a given $x \in A$. Even with reduced bounds, cases may arise where $n_x$ does not reduce at all. For example, if $m=10, m'=5, l=30 \text{ and } l'=20$, then the loop can still run for, say 15, iterations both before and after condition strengthening. Thus, a quantity of greater significance is the difference in these two relaxations, which will then represent the number of iterations surely cut down due to condition strengthening. Let us denote this quantity by $\sigma_{C_{A}' \vdash C_{A}}(f)$ and define it by the following:
	
	\begin{equation}
	\label{eq:sigma_stren}
	\sigma_{C_{A}' \vdash C_{A}}(f) = \Delta \hat{\Theta}_{C_{A},C_{A}'} (f) - \Delta \Theta_{C_{A},C_{A}'} (f)
	\end{equation}
	
	The value of $\sigma_{C_{A}' \vdash C_{A}}(f)$ can be negative, zero or positive, all three with different consequences. A negative value indicates that the relaxation in order is higher than the relaxation in limit and hence, a high negative value is an indicator of a possibility of relatively earlier loop termination. A positive value of $\sigma_{C_{A}' \vdash C_{A}}(f)$ indicates otherwise. It means that the relaxation in order is lower than the relaxation in limit, which is good for the worst case termination of the loop, but does not help much in the best, or even the average case. A zero value, however, shifts the entire iteration count window to the left. A particularly interesting case here is when the limit is relaxed by more than the original window of the loop iteration count. In this situation, we have $l' < m$ and this means that strengthening the condition has significantly lowered the number of times the loop will execute, because now, the upper bound on this iteration count is smaller than the original lower bound on this count. Consequently, for all $x \in [A]_{C_A'}$, the values of $n_x$ are guaranteed to have reduced significantly. For example, consider the situation when $C_{A}$ is strengthened to the maximum extent possible, i.e. $C_{A}'=\textit{False}$. In this situation, the set $[A]_{C_{A}'}=\phi$, the empty set, and hence, the loop does not execute at all for any $x \in A$.\\
	
	We can visualize this relation of condition strengthening with the lowering of loop iteration count through the formulation of an order-preserving monotonic function. We know that the specification of any Boolean condition $C \in \mathcal{C}_A$ for selection of elements in $A$ is equivalent to selecting a subset of $A$, in which all elements satisfy $C$ and no element outside this subset satisfies $C$. This subset was denoted previously by $[A]_C$. Clearly, we have $[A]_C \subseteq [A]$, and thus, $\left| [A]_C \right| \leq |A|$. This holds true even if $A$ is an infinite set. Thus, if we have two Boolean conditions $C_1, C_2 \in \mathcal{C}_A$, such that $C_2 \vdash C_1$, then the number of elements in $A$ satisfying $C_2$ cannot be more than the number of elements satisfying $C_1$, and hence,
	\begin{equation}
	C2 \vdash C1 \quad \implies \quad \left| [A]_{C_2} \right| \leq \left| [A]_{C_1} \right|
	\end{equation}
	Consequently, any element in $[A]_{C_2}$ cannot satisfy $C_2$ under repeated iteration of any $f:A \to A$ more than the number of times it will satisfy $C_1$. Hence, we also have,
	\begin{equation}
	C2 \vdash C1 \quad \implies \quad n_{x,C_2} \leq_f n_{x,C_1}
	\end{equation}
	where $n_{x,C}$ denotes the order of $x$ with respect to the truth preservation of $C$ by some function $f$. This implication is true only when the truth preserving function under consideration is $f$ for the calculation of both $n_{x,C_1}$ and $n_{x,C_2}$, hence, the subscript $f$ under $\leq$ sign. Now, let $N_f:\mathcal{C}_A \to \mathbb{Z}^+ \times \mathbb{Z}^+$ be a function which returns, for a given Boolean condition $C$ and a function $f$, a tuple containing the order and limit of $f$'s truth preservation of $C$, i.e.
	\begin{equation}
	N_f(C) = \left( \Theta_C(f), \hat{\Theta}_C(f) \right), \quad \forall C \in \mathcal{C}_A \text{ and } f:A \to A
	\end{equation}
	Let the tuples in $(\mathbb{Z}^+ \times \mathbb{Z}^+)$ be lexicographically ordered by setting $(a,b) < (a',b')$ if either $a < a'$ or ($a = a' \text{ and }b < b'$). We have $(a,b) = (a',b')$ only when $a = a'$ and $b=b'$. Now, for a given $f$, it can be said that
	\begin{equation}
	C2 \vdash C1 \quad \implies \quad N_f(C_2) \leq N_f(C_1), \quad \forall C_1,C_2 \in \mathcal{C}_A
	\end{equation}
	The elements of $\mathcal{C}_A$ can also be ordered under standard ordering for Boolean conditions, where we say that $C_2 <_\mathcal{B} C_1$, if $C_2 \vdash C_1$ and $[A]_{C_2} \subset [A]_{C_1}$. The case $C_2 =_\mathcal{B} C_1$ arises only when $[A]_{C_2} = [A]_{C_1}$. Hence, we can now visualize $N_f$ as an order-preserving function between the posets $\left( \mathcal{C}_A, \leq_\mathcal{B} \right)$ and $\left( \mathbb{Z}^+ \times \mathbb{Z}^+, \leq_f \right)$.
	
	\subparagraph*{Proof of undecidability of $N_f(C)$ :} The problem of computing $N_f(C)$ can be shown to be undecidable, in general, by reducing the instances of Halting problem to instances of determining $N_f(C)$ for some computable function, $f$ and an appropriate Boolean condition, $C$. We do this by observing that for any $C \in \mathcal{C}_A$, if an algorithm to compute $N_f(C)$ had existed, then it would have been possible to extract two numbers $m,n \in \mathbb{Z}^+$, such that $N_f(C)=(m,n)$ for any given function $f$. This would mean that we can deterministically compute an upper bound on the number of times a loop with condition $C$ would execute. This is the same as asking if this upper bound is finite and in case it is, what its face value is. In other words, we are trying to ask the question whether the loop in the code snippet as in Fig. \ref{Fig:loop} ever stops its execution or if the code produces any output for any $x \in A$. Any algorithm that can answer this question for this loop can surely solve the general Halting problem as well. Hence, the task of solving the Halting problem reduces to the task of computing $N_f(C)$, which renders the latter undecidable. 
	
	\subsubsection{Condition Weakening} The case of weakening $C_{A}$ to $C_{A}'$ is similar to condition strengthening with only minor differences. We now have $C_{A} \vdash C_{A}'$, and hence, the falsehood of $C_{A}'$ surely implies the falsehood of $C_{A}$. However, when $C_{A}$ becomes false, $C_{A}'$ may still be true. Thus, the new order and limit are both increased. 
	\begin{subequations}
	\begin{align}
	\Theta_{C_{A}} (f) \leq \Theta_{C_{A}'} (f)\\
	\hat{\Theta}_{C_{A}} (f) \leq \hat{\Theta}_{C_{A}'} (f)
	\end{align}
	\end{subequations}
	
	The definitions of order and limit relaxations remain unchanged because we are interested in the change of order and limit with respect to the change in condition. Hence, in this case, the two relaxations will be negative, indicating an increase in the values. The value of $\sigma_{C_{A} \vdash C_{A}'}(f)$ is then interpreted accordingly. The case of special interest here is when $C_{A}$ is weakened to the maximum extent, i.e. $C_{A}'=\textit{True}$. In this situation, we have $[A]_{C_{A}'}=A$ and the condition is true for all $x \in 
	A$. Thus, $f_{C_{A}'}$ becomes an infinite order truth preserving function and the loop executes \textit{ad infinitum} everytime.\\
	
	For the case of condition weakening, the function $N_f(C)$ is anti-monotonic instead. A weaker condition yields larger (or atleast similar) values of order and limit, and hence, we have
	\begin{equation}
	C_2 \leq_\mathcal{B} C_1 \quad \implies \quad N_f(C_2) \geq_f N_f(C_1), \quad \forall C_1,C_2 \in \mathcal{C}_A
	\end{equation} 
	
	\paragraph*{Nesting of Loops : } We now look at an interesting example to better understand the effect of condition strengthening and weakening on orders and limits. Let $f:A \times B \to A \times B$ be a function and $C_A \in \mathcal{C}_A$, $C_B \in \mathcal{C}_B$ be two Boolean conditions. Now, consider the code snippet as in Fig. \ref{Fig:NestedLoop}. It has been written as an extension to what we saw in Fig. \ref{Fig:loop}. \\
	
	\begin{figure}[htp]
	\texttt{
	\begin{algorithmic}
	\State Initialize $x \in A$
	\Comment Outer code C1 starts
	\While{$C_{A}$ is \emph{True} on $x$}
		\State Initialize $y \in B$
		\Comment Inner code C2 starts
		\While{$C_{B}$ is \emph{True} on $y$}
			\State Set $(x,y) \gets f(x,y)$
		\EndWhile
		\State Print $y$
		\Comment Inner code C2 ends
	\EndWhile
	\State Print $x$
	\Comment Outer code C1 ends
	\end{algorithmic}
	}
	\caption{Example of nested loops} \label{Fig:NestedLoop}
	\end{figure} 
	
	 The inner code C2 is exactly the same as the one in Fig. \ref{Fig:loop}. If we ignore C1 for the moment, the computation of C2 is independent of the value $x$ takes. Thus, $x$ can be chosen without any restrictions and $C_A$ can be interpreted as being $\emph{True}$ in this case. Now, when C1 enters into the scene, $C_A$ is strengthened and we choose only some values of $x$ for computation of $f$. In a way, we have changed our looping conditions in a manner so that only a fewer tuples $(x,y)$ will undergo trasformation under $f$. Thus, the number of times we print $y$ will reduce as compared to the situation where $C_A = \emph{True}$. The number of times we print $x$ is surely not less than what it would have been had $C_B = \emph{True}$. We can, thus, safely conclude that nesting of loops is equivalent to restricting the number of times any individual loop would execute and is, therefore, equivalent to condition strengthening. The stronger condition in this case is $C_A(x) \land C_B(y)$, for which $\Theta_{C_A(x) \land C_B(y)}(f) \leq \Theta_{C_B(y)}(f)$ and $\hat{\Theta}_{C_A(x) \land C_B(y)}(f) \leq \hat{\Theta}_{C_B(y)}(f)$.\\
	
	Let us now consider a different type of nesting that may arise. This time, the iteration count of the inner loop is increased indefinitely by enveloping it with an infinite loop, as shown in Fig.  \ref{Fig:NestedLoop2}. \\
	
	\begin{figure}[htp]
	\texttt{
	\begin{algorithmic}
	\While{\emph{True}}
		\State Initialize $x \in A$
		\While{$C$ is \emph{True} on $x$}
			\State Set $x \gets f(x)$
		\EndWhile
		\State Print $x$
	\EndWhile
	\State Print $"Done"$
	\end{algorithmic}
	}
	\caption{Another example of nested loops} \label{Fig:NestedLoop2}
	\end{figure} 
	
	The outer loop iterates indefinitely in the absence of any stopping condition. However, in one iteration of the outer loop, the execution of the inner loop is exactly equivalent to what it would have been in the absence of the outer loop. Also, the iteration of inner loop in one iteration of the outer loop is independent of all others because the variable $x$ is reinitialized everytime. Thus, the order and limit of the inner loop do not change. Even if we treat this example as a special case of the example as in Fig. \ref{Fig:NestedLoop}, the stronger condition, here, will be $\emph{True} \land C$, which is clearly equivalent to $C$ itself. Hence, we have really \emph{not} strengthened our condition for the iteration of inner loop. The order and limit must not change.\\
	
	It is worth talking a little about the order and limit of the outer loop in Fig. \ref{Fig:NestedLoop2} as well. We are sure that the loop executes indefinitely, but perhaps the underlying function $g$ which preserves the truth of this loop condition $\emph{True}$ is not obvious. If we try to rewrite the code, focussing primarily on the outer loop, we get something similar to Fig. \ref{Fig:NestedLoop3}.
	
	\begin{figure}[htp]
	\texttt{
	\begin{algorithmic}
	\State Initialize $v \in V$
	\Comment $V : $ Set of state variables 
	\While{\emph{True}}
		\State Set $v \gets g(v)$
	\EndWhile
	\State Print $"Done"$
	\end{algorithmic}
	}
	\caption{Analyzing the outer loop of Fig. \ref{Fig:NestedLoop2}} \label{Fig:NestedLoop3}
	\end{figure} 
	
	As it can be observed, we have considered a variable $v$, not present in Fig. \ref{Fig:NestedLoop2}, to belong to the set of state variables. This is done to illustrate the fact that the function $g$ is bound to have a definite domain and we cannot just leave it undefined. The set $V$ is an abstraction of any and everything that may be thought of as constituting the state variables for our code, with the condition that $x$ must lie in this set for sure. This way, any modification done to $x$ by the inner loop is reflected as a modification in $v$ through $g$. We can express $g$ as a partial function in terms of $f$, the steps for which are provided in later sections. For now, assuming that $g$ mimics the behavior of the body of this outer while-loop, we can now say that $\Theta_{\emph{True}}(g) = \hat{\Theta}_{\emph{True}}(g) = \infty$. 
	 
	\subsubsection{Smaller state variable set} The situation arises when we replace $A$ with some $A' \subset A$ so that the number of $x$'s in $A'$ that satisfy $C_{A}$ may now be smaller. This is equivalent to replacing $A$ with $[A]_{C'}$ for some Boolean condition $C' \in \mathcal{C}_A$ defined on the elements of $A$ such that $A'=[A]_{C'}$. The code in Fig. \ref{Fig:smallStateVar} mimics this situation.
	
	\begin{figure}[htp]
	\texttt{
	\begin{algorithmic}
	\State Initialize $x \in A$
	\While{$C'$ is False on $x$}
	\Comment{Allows only $x \in [A]_{C'}$}
	\EndWhile
	\While{$C_{A}$ is True on $x$}
	\Comment{Main loop body}
		\State Set $x \gets f(x)$
	\EndWhile
	\State Print $x$
	\end{algorithmic}
	}
	\caption{Example for restricting the set of state variables} \label{Fig:smallStateVar}
	\end{figure} 
	
	The output of the above code is seen for only those values of $x$ that satisfy $C'$ and are able to render $C_{A}$ false through zero or a finite number of iterations of $f$. This is similar to restricting the initialization of $x$ to $[A]_{C'}$ instead of $A$. We will then need the order and limit of $f$ with respect to $C_{A}$, but on the elements of $[A]_{C'}$ this time. A simpler way to verify this is through the code in Fig. \ref{Fig:smallStateVar2}, which is equivalent to that in Fig. \ref{Fig:smallStateVar}.\\
	
	\begin{figure}[htp]
	\texttt{
	\begin{algorithmic}
	\State Initialize $x \in A$
	\While{$(C_{A} \land C')$ is True on $x$}
	\Comment{Loop condition changed}
		\State Set $x \gets f(x)$
	\EndWhile
	\While{$C'$ is False on $x$}
	\Comment{To print only if $x \in [A]_{C'}$}
	\EndWhile
	\State Print $x$
	\end{algorithmic}
	}
	\caption{Rewritten code for Fig. \ref{Fig:smallStateVar}} \label{Fig:smallStateVar2}
	\end{figure} 
	
	The equivalence of the codes in Fig. \ref{Fig:smallStateVar} and Fig. \ref{Fig:smallStateVar2} is based on the computation performed on different values of $x$ and the final output produced. It can be said with absolute surity that the outputs in these two cases is exactly the same and the two programs halt under exactly the same conditions. Consequently, the sets of all $x$ for which the programs do not halt are the same in the two cases. Thus, if we now want to analyze the effect on the iteration count of out main loop, Fig. \ref{Fig:smallStateVar2} suggests that this is precisely the case of condition strengthening from $C_A$ to $C_A \land C'$ so that the new truth preservation order and limit for $f$ are not larger than the original values. 
	 
	\subsubsection{Larger state variable set} The case of a larger state variable set means that we are now initializing $x$ from a set $A'$, where $A \subset A'$, so that the number of elements in $A'$ that satisfy $C_A$ may increase. A decrease in this number is surely not possible, so that possibility can be ruled out trivially. The set $A'$ is called \emph{larger} keeping in mind the possible increase in the number of $x$ satisfying $C_A$. The notion has nothing to do with the cardinalities of $A$ and $A'$, which may be the same for infinite sets. Consider the set $A' \backslash A$. For any $y \in A'\backslash A$, we cannot be sure if $C_A$ is defined on $y$ because all we know about $C_A$ is that it is defined on elements in $A$. The extension of $A$ to $A'$ may render $C_A$ undefined on elements in $A' \backslash A$. For this, let $\mathcal{C}_{A'}$ be the analog of $\mathcal{C}_A$ for $A'$, i.e. the set of all Boolean conditions defined on the elements of $A'$. Define $C' \in \mathcal{C}_{A'}$ such that
	\begin{equation}
	C_A(x) = C'(x) \quad \forall x \in A
	\end{equation} 
	This way, for every $x \in A$, the truth of $C_A$ is preserved by $C'$. Note that we did not use a logical implication in place of logical equivalence here because we do not want the state of $C'(x)$ to be \emph{True} when $C_A(x)$ is \emph{False}. With this formulation, we can be sure that the order of $f$ with respect to the truth preservation of $C'$ remains unchanged, atleast for the elements in $A$. The new order and limit of $f$ are given in Eq. \ref{Eq:largerStateVar}.
	
	\begin{subequations}
	\label{Eq:largerStateVar}
	\begin{align}
	\Theta_{C'}(f) = \min \left( \Theta_{C_A}(f), \min_{x \in [A'\backslash A]_{C'}} \Theta_{f,C'}(x)  \right) \leq \Theta_{C_A}(f)\\
	\hat{\Theta}_{C'}(f) = \max \left( \hat{\Theta}_{C_A}(f), \max_{x \in [A'\backslash A]_{C'}} \hat{\Theta}_{f,C'}(x) \right) \geq \hat{\Theta}_{C_A}(f)
	\end{align}
	\end{subequations}
	
	Thus, we see that unlike previous cases, where both order and limit were either decreasing or increasing, we only see a possible decrease in the order here. The limit is likely to increase, but can never be smaller than the original value. The function $N_f(C')$ is neither monotonic nor anti-monotonic in this case. 
	
	\subsubsection{Different state variable set with equal or larger cardinality} The case of changing the state variables entirely and bringing in a newer set can have some very interesting consequences. Everything now depends on how the previously used state variables map into the new ones. We also need to take care of how the Boolean conditions map from one set to another. The point will be expanded upon in the discussion below. \\
	
	Let us first analyze the case where the new set, say $B$, has equal or larger cardinality than the original set, say $A$, of state variables. In this case, there will always a subset, $B' \subseteq B$, of the new state variable set which will be isomorphic to the original state variable set. By isomorphism, we mean that $Dom(\varphi)=B'$ and $|A| = |B'|$ such that $\varphi:A \to B'$ is a bijection. Thus, corresponding to every condition $C_A \in \mathcal{C}_A$, we can specify the corresponding condition, $C \in \mathcal{C}_B$ for elements in $B$ such that $[B]_C=\varphi([A]_{C_A})=B'$. This can also be written as 
	
	\begin{equation}
	\label{Eq:mappingCtoDiffSet}
	C = \left[ y \in B \mid C_A(x) \land y = \varphi(x), x \in A \right]
	\end{equation}
	
	However, $C$ is not necessarily the condition that will be given to us for elements in $B$. In other words, the condition $C_B \in \mathcal{C}_B$ which is specified in terms of the new state variables may not be the same as $C$. We may have either of the three cases true : $[B]_C = [B]_{C_B}$, $[B]_C \subset [B]_{C_B}$ and $[B]_{C_B} \subset [B]_C$. The truth preservation of $f$ will now have to be investigated for $C_B$.  To be precise, our function $f$ is now changed to $(\varphi \circ f)$ to account for the change in the state variable set. Let us deal with these three cases individually. 
	
	\subparagraph*{Case I : $[B]_C = [B]_{C_B}$}
	This case arises when $C =_\mathcal{B} C_B$. For this situation, we can take advantage of our bijective mapping $\varphi$ to conclude that the truth preservation order and the corresponding limit do not change.
	
	\begin{subequations}
	\begin{align}
	\Theta_{C_B}(\varphi \circ f) = \Theta_{C}(\varphi \circ f) = \Theta_{C_A}(f) \\
	\hat{\Theta}_{C_B}(\varphi \circ f) = \hat{\Theta}_{C}(\varphi \circ f) = \hat{\Theta}_{C_A}(f)
	\end{align}
	\end{subequations}
	
	\subparagraph*{Case II : $[B]_C \subset [B]_{C_B}$} This case arises when $C <_\mathcal{B} C_B$. For this situation, we use the rules for condition weakening, to conclude that the new truth preservation order and the corresponding limit are not less than the original values.
	
	\begin{subequations}
	\begin{align}
	\Theta_{C_B}(\varphi \circ f) \geq \Theta_{C}(\varphi \circ f) = \Theta_{C_A}(f) \\
	\hat{\Theta}_{C_B}(\varphi \circ f) \geq \hat{\Theta}_{C}(\varphi \circ f) = \hat{\Theta}_{C_A}(f)
	\end{align}
	\end{subequations}
	
	\subparagraph*{Case III : $[B]_{C_B} \subset [B]_C$} Similar to the previous case, this situation arises when $C_B <_\mathcal{B} C$. Thus, we use the rules for condition strengthening, to conclude that the new truth preservation order and the corresponding limit are not more than the original values.
	
	\begin{subequations}
	\begin{align}
	\Theta_{C_B}(\varphi \circ f) \leq \Theta_{C}(\varphi \circ f) = \Theta_{C_A}(f) \\
	\hat{\Theta}_{C_B}(\varphi \circ f) \leq \hat{\Theta}_{C}(\varphi \circ f) = \hat{\Theta}_{C_A}(f)
	\end{align}
	\end{subequations}
	
	Note that throughout this discussion, we are not talking of a new function defined on the elements of $B$ for its truth preservation of $C_B$. The function of interest is $(\varphi \circ f)$, for which the domain is $A$.
	
	\subsubsection{Different state variable set with smaller cardinality} 
	
	 Next, we discuss the case of mapping our original state variables into a smaller set. As will be evident, this turns out to be more complex than the case first appears to be. Following similar notations as above, let $A$ be the old set of state variables, which is modified into the new set $B$ through a mapping $\varphi:A \to B$. Since we know that the cardinality of $B$ is strictly smaller than that of $A$, we can be sure that $\varphi$ is not an injective map. It need not be surjective as well, as there is no formal restriction to map into every element in $B$. Thus, two cases arise here - one where $\varphi$ is surjective and the other where it is not. We will deal with these two cases one by one.
	
	\subparagraph*{Case I : $\varphi : A \to B$ is surjective }When the state-variable mapping function is known to be surjective, or onto, we can be sure that every element in $B$ has a pre-image in $A$. The problem lies in the fact that this pre-image is not unique. Since the cardinality of $B$ is strictly less than that of $A$, there must exist atleast two elements in $A$ which map to the same element in $B$. For any $A' \subset A$, let $\varphi(A') = \{ y \in B : \varphi(x)=y \text{ for some }x \in A' \}$, i.e. the union of the images of all elements in $A'$. Similarly, for any $y \in B$, let $\varphi^{-1}(y) = \{ x \in A : \varphi(x)=y \}$, i.e. the set of all elements in $A$ that map to $y$. Now, given a condition $C_A \in \mathcal{C}_A$, form a Boolean condition $C \in \mathcal{C}_B$ similar to Eq. \ref{Eq:mappingCtoDiffSet}. Unlike the previous situation, this time the set $[B]_C$ is such that $[A]_{C_A} \subseteq \varphi^{-1}([B]_C)$ because it is not necessary that the elements in $[B]_C$ inverse map only into elements of $[A]_{C_A}$. Thus, the truth preservation of $C$ by $(\varphi \circ f)$ can be treated as a case of condition weakening. However, the elements in $\varphi^{-1}(y)$ that do not satisfy $C_A$ will not contribute to a change in the order and limit of $f$. Hence, we have
	
	\begin{subequations}
	\begin{align}
	\Theta_{C}(\varphi \circ f) = \Theta_{C_A}(f)\\
	\hat{\Theta}_{C}(\varphi \circ f) = \hat{\Theta}_{C_A}(f)
	\end{align}
	\end{subequations}
	
	Now, for a given $C_B \in \mathcal{C}_B$, we can form cases as what we did while mapping into a different, larger state variable set and obtain results exactly similar to the ones obtained then.
	
	\subparagraph*{Case II : $\varphi : A \to B$ is not surjective } This case arises when not all elements in $B$ have a preimage in $A$, but there certainly exists some subset $B' \subset B$ for which $\varphi:A \to B'$ is surjective. Thus, given some $C_B \in \mathcal{(C)_B}$, we can first use Case I for mapping into $B'$ and then use the results of larger state variable set to account for those elements in $B \backslash B'$ which satisfy $C_B$.
	
	\subsubsection{Using conditional functions}
	
	Computer programs often employ conditional execution of functions based on the truth of some given Boolean condition on the state variables. The kind of functions that we will deal with now have the general form
	\begin{equation}
	\label{eq:condFunc}
	f(x) = 
	\begin{cases}
	f_1(x) & \text{if $C(x)$ is \emph{True}} \\
	f_2(x) & \text{otherwise}
	\end{cases}
	\end{equation}
	
	In all our analysis so far, the execution of $f(x)$ was limited to $x \in [A]_C$ at best, and left undefined otherwise. As will be evident later, this forms a special case of Eq. \ref{eq:condFunc}. The code snippet that mimics the behavior of this function inside a \texttt{while} loop conditioned on $C_A$ is given in Fig. \ref{Fig:condFunc}. 
	
	\begin{figure}[htp]
	\texttt{
	\begin{algorithmic}
	\State Initialize $x \in A$
	\While{$C_{A}(x)$ is \emph{True}}
		\If{$C(x)$ is \emph{True}}
		\Comment Conditional function
			\State Set $x \gets f_1(x)$
		\Else
			\State Set $x \gets f_2(x)$
		\EndIf
	\EndWhile
	\State Print $x$
	\end{algorithmic}
	}
	\caption{Conditional function inside a \texttt{while}-loop} \label{Fig:condFunc}
	\end{figure} 
	
	To study the behavior of $f$ with respect to its truth preservation order and limit of $C_A$, we need some extra information this time. Through a simple unrolling operation on this loop, we see that $f_1(x)$ is executed when both $C_A(x)$ and $C(x)$ are \emph{True}. Similarly, $f_2(x)$ is executed when $C_A(x)$ is \emph{True} but $C(x)$ is \emph{False}. Thus, we need the truth preservation orders and limits of $f_1$ and $f_2$ with respect to $(C_A \land C)$ and $(C_A \land \lnot C)$, respectively. This can be verified by convincing ourselves that the code in Fig. \ref{Fig:condFunc} is equivalent to the one in Fig. \ref{Fig:condFunc2}.
	
	\begin{figure}[htp]
	\texttt{
	\begin{algorithmic}
	\State Initialize $x \in A$
	\While{$C_{A}(x)$ is \emph{True}}
		\While{$C_A(x) \land C(x)$ is \emph{True}}
			\State Set $x \gets f_1(x)$
		\EndWhile
		\While{$C_A(x) \land \lnot C(x)$ is \emph{True}}
				\State Set $x \gets f_2(x)$
			\EndWhile
	\EndWhile
	\State Print $x$
	\end{algorithmic}
	}
	\caption{Modified code for a conditional function inside a \texttt{while}-loop} \label{Fig:condFunc2}
	\end{figure}
	
	Let us assume the following values to be known, before we proceed with further analysis.
	
	\begin{equation*}
	\Theta_{(C \land C_A)}(f_1)=m_1' , \quad \Theta_{(\lnot C \land C_A)}(f_2)=m_2', 
	\end{equation*}
	\begin{equation*}
	\hat{\Theta}_{(C \land C_A)}(f_1)=l_1', \quad \hat{\Theta}_{(\lnot C \land C_A)}(f_2)=l_2'
	\end{equation*}
	
	The conditions for execution of either $f_1$ or $f_2$ guarantee that if $C_A(x)$ is \emph{True}, then one of these functions must execute. Also, $\Theta_{C \land C_A}(f_1)=m_1'$ tells us that there exists at least one $x \in [A]_{C \land C_A}$ for which $\Theta_{C \land C_A,f_1}(x)=m_1'$ and for all $y \in [A]_{C \land C_A} \backslash \{ x \}$, we have $\Theta_{C \land C_A,f_1}(y) \geq m_1'$. Hence, Fig. \ref{Fig:condFunc2} helps us conclude that 
	
	\begin{equation}
	\Theta_{C_A}(f) \geq \min_{x \in [A]_{C_A}} \left( \Theta_{C \land C_A, f_1}(x) + \Theta_{\lnot C \land C_A,f_2}
	 \left( f_1^{(m_1'+1)}(x) \right) + 1 \right)
	\end{equation}
	
	The extra factor of one is added to account for the termination of first \texttt{while}-loop. The code in Fig.\ref{Fig:condFunc2} can be rewritten by exchanging the order of two \texttt{while}-loops, without modifying the output. In that case, we will have
	
	\begin{equation}
	\Theta_{C_A}(f) \geq \min_{x \in [A]_{C_A}} \left( \Theta_{\lnot C \land C_A, f_2}(x) + \Theta_{C \land C_A,f_1}
	 \left( f_2^{(m_2'+1)}(x) \right) + 1 \right)
	\end{equation} 
	
	In both cases, we use the linearity of \emph{min} operator over addition to obtain
	
	\begin{subequations}
	\begin{align}
	\Theta_{C_A}(f) \geq 1 + m_1' + \min_{x \in [A]_{C_A}} \left( \Theta_{\lnot C \land C_A,f_2}
	 \left( f_1^{(m_1'+1)}(x) \right) \right)\\
	\Theta_{C_A}(f) \geq 1 + m_2' + \min_{x \in [A]_{C_A}} \left( \Theta_{C \land C_A,f_1}
	 \left( f_2^{(m_2'+1)}(x) \right) \right)
	\end{align}
	\end{subequations}
	
	In each of these equations, the last term added is a minima over the number of times second loop executes for all $x \in [A]_{C_A}$ after the first \texttt{while}-loop has terminated. This depends entirely on the nature of $f_1$ and $f_2$, and no definite bounds can be established as the moment. However, we do know that the value of this minima will either be $m_1'$ or $-1$ for $f_1$ and $m_2'$ or $-1$ for $f_2$. Thus, we can say with certainty that
	
	\begin{equation}
	\Theta_{C_A}(f) \geq \min \left( m_1', m_2', 1+m_1'+m_2' \right) = \min \left( m_1', m_2' \right)
	\end{equation}
	
	Clearly, this lower bound on the truth preservation order of $f$ is very weak, and this prevents us from expressing the inequality above into an equality. Through similar arguments, we can prove that 
	
	\begin{equation}
	\hat{\Theta}_{C_A}(f) \geq \min \left( l_1', l_2' \right)
	\end{equation}
	
	Thus, for the case when we have a conditional function inside a \texttt{while}-loop, the effective truth preservation order and limit are hard to determine if the nature of this conditional function is unknown. However, given some $x \in [A]_{C_A}$, the exact value of its order can be found out through the following recursive relation.
	
	\begin{equation}
	\begin{split}
	\Theta_{C_A,f}(x) = \underbrace{\Theta_{C \land C_A,f_1}(x)}_\text{m} + \underbrace{\Theta_{\lnot C \land C_A, f_2}\left( f_1^{(m+1)}(x) \right)}_\text{n} + \\ \Theta_{C_A,f}\left( f_2^{(n+1)} \circ f_1^{(m+1)}(x) \right)
	\end{split}
	\end{equation}
	
	A lot of other factors can affect the values of truth preservation order and limit of a given function which makes it impossible to enumerate them all. We, thus, save further analysis in this domain for later, when the need arises. 
	
	\section{A special symbol : $\perp$}
	
	Dealing with partial functions leaves one important question unanswered. If the function $f$ is defined for only elements in its domain, then how would one explain its behavior on the remaining elements if the need to generalize ever arises. Consider Eq. \ref{Eq:idc} for example. The analysis following this equation tells us that we can model the behavior of $id_C$ through the following code.
	
	\begin{figure}[htp]
	\texttt{
	\begin{algorithmic}
	\State Initialize $x \in A$
	\While{$C(x)$ is \emph{False}}
	\EndWhile
	\State Print $x$
	\end{algorithmic}
	}
	\caption{Code for computing $id_C$}
	 \label{Fig:idcCode}
	\end{figure}
	
	Any computer program containing this code will enter into a non-terminating computation for all $x \not\in [A]_C$. This can be modelled by visualizing $id_C$ as a function which operates on large but finite collection, $\mathcal{S}$, of all possible valid assignments of state-variables (including $x$) such that every application of $id_C$ allows us to jump from one valid assignmentm $S_1 \in \mathcal{S}$, to another valid assignment, $S_2 \in \mathcal{S}$. Infact, all partial functions can be assumed to be working this way, the difference from the total functions being that while the latter can operate on all $S \in \mathcal{S}$, the former only modifies a sub-collection, $\mathcal{S}' \subset \mathcal{S}$. With this picture in mind, we need some element of $\mathcal{S}$ to represent invalid assignments for all state variables, for which we lift this collection to obtain $\mathcal{S}_\perp = \mathcal{S}\cup \{ \perp \}$. Now, we have an assignment $\perp$ of the state variables which can be interpreted as being an \emph{undefined} state, since we do not know how the state variables, or any function, would behave on reaching here. Thus, we can safely say that this state acts as a trap in the machine so that entering this state will never allow us to come out of it. It, then, becomes easy to define our partial function, $id_C$ as
	
	\begin{equation}
	id_C(x) = 
	\begin{cases}
	x & \text{if $x \in [A]_C$} \\
	\perp & \text{otherwise}
	\end{cases}
	\end{equation}
	
	The signature of $id_C$ now becomes $\mathcal{S}_\perp \to \mathcal{S}_\perp$. Similarly, all partial functions can be defined on their lifted domains and ranges as above. In some cases, we may also need to define a special function, called the \emph{Undefined function}, which maps every element in $\mathcal{S}_\perp$ to $\perp$. Let this function be denoted by $id_\perp$. Hence, $id_C$ can also be written, similar to conditional functions, as
	
	\begin{equation}
	id_C(x) = 
	\begin{cases}
	id(x) & \text{if $x \in [A]_C$} \\
	id_\perp(x) & \text{otherwise}
	\end{cases}
	\end{equation}
	
	The truth preservation order of $id_\perp$ with respect to any condition $C \in \mathcal{C_\mathcal{S}}$ can be assumed $-1$ because the state variables in this undefined state are unable to satisfy any condition due to the invalidity of their values. More precisely, the unsatisfaction of $C$ does not imply the falsehood of $C$, rather the inapplicability of $C$ over the state variables in this set. Similarly, composition of $id_\perp$ with any function can be interpreted to produce $id_\perp$ as the output, because of the trapping nature of $\perp$. Thus, we have $(f \circ id_\perp) = (id_\perp \circ f) = id_\perp$ for all partial functions $f:\mathcal{S}_\perp \to \mathcal{S}_\perp$.\\
	
	The notion of \emph{undefinability} becomes vague and difficult to handle very easily. For example, adding a constant to all state variables in the undefined state leaves us in this state only, since we cannot be sure what result would come out of such an operation. If, then, we interpret $\perp$ as a state of no information, a special operator, say $\bullet$, can be used to denote the partial information gained on performing some operation on $\perp$.  In other words, we can say that 
	
	\begin{equation}
	x \bullet \perp = \perp \bullet x = x \quad \forall x \in \mathcal{S}
	\end{equation} 
	
	For example, in the case of conditional functions of the form Eq.  \ref{eq:condFunc}, we can write $f = (f_1 \circ id_C) \bullet (f_2 \circ id_{\lnot C})$. This way, for $y \in [A]_C$, we have, 
	
	\begin{equation*}
	\begin{split}
	f(y) & = ((f_1 \circ id_C) \bullet (f_2 \circ id_{\lnot C}))(y)\\
	& = (f_1 \circ id_C)(y) \bullet (f_2 \circ id_{\lnot C})(y)\\
	& = (f_1 \circ id_C)(y) \bullet (f_2 \circ id_\perp)(y)\\
	& = f_1(y) \bullet \perp\\
	& = f_1(y)
	\end{split}
	\end{equation*}
	
	Now, $f(x)$ can be interpreted as providing whatever partial information is known about $y$ through the application of either $f_1$ or $f_2$, conditioned on $C$. Note that $\bullet$ is \emph{not} a way to come out of the undefined state. We still have $\perp \bullet \perp = \perp$ to be true and hence, a non-terminating behavior on reaching $\perp$ will still be seen.
	
	\section{Infinite order truth preserving functions}
	
	A special class of truth preserving functions is the class of absolutely truth preserving functions, in which every function has an infinite order with respect to a given condition $C$. The simplest example, perhaps, is for the set $A = \{ 0,1,2,3,\dots \}$, where $C = [i > 0]$ and $f(x) = x+1$. For every $x \in [A]_C$, $f(x) \in A$ satisfies $C$, and hence, the truth preservation order for $f$ is infinite. Needless to say, the limit of an absolutely truth preservaing function is infinite as well. \\
	
	\begin{figure}[htp]
	\texttt{
	\begin{algorithmic}
	\State Initialize $x \in A$
	\While{$C(x)$ is \emph{True}}
		\State Set $x \gets f(x)$
	\EndWhile
	\State Print $x$
	\end{algorithmic}
	}
	\caption{Code to demonstrate infinite order of truth preservation}
	 \label{Fig:InfOrder}
	\end{figure}
	
	The study of functions exhibiting such behavior is particularly interesting because it is precisely these functions that are responsible for non-terminating behaviour arising in our programs. Consider the code snippet in Fig. \ref{Fig:InfOrder}, assuming that $f:[A]_C \to [A]_C$ has an infinite order of truth preservation for $C$. The case of indefinite computation is unavoidable in this situation. If we try to generalize the conditions under which functions like $f$ can have an infinite order, we can conclude that it can be due to infiniteness (countable or uncountable) of the domain and range or the existence of $k$-periodic points of $f$. Let us study these cases in detail.
	
	\paragraph*{Existence of fixed-points of $f$}
	For a given function $f:A \to A$, we say that $x \in A$ is a fixed point of $f$ if it satisfies the condition $f(x)=x$, i.e. it is invariant under the application of $f$. This means $f^{(k)}(x)=x$ for all $k \geq 0$, and hence, the orbit of $x$ under $f$ contains the single point $\{x\}$. Let us consider the orbit for any given $x$ to be an ordered set such that $x \preceq f(x) \preceq f(f(x)) \preceq \dots$. This way, if for any $y \in A$, we have $y \preceq (f^{(k)}(y) = y)$ for some $k \geq 1$, we can conclude that the orbit of $y$ under $f$ is periodic with period $k$. We discuss only the special case, $k=1$ here, leaving the generalized analysis for later. \\
	
	A function can have more than one fixed points. Let us denote the set of fixed points of $f$ by $\texttt{fix}(f)$. The case of infinite order truth preservation arises in one of the following two situations:
	\begin{enumerate}
	\item We start our computation for some $x \in \texttt{fix}(f) \cap [A]_C$.
	\item We start our computation for some $x \in [A]_C$ and eventually reach some $y \in \texttt{fix}(f)$ such that $y \in [A]_C$. 
	\end{enumerate} 
	
	The first case is trivial to explain, since starting at any $x \in \texttt{fix}(f)\cap [A]_C$ will always produce $f(x)=x$ and hence, the condition $C$ will be \emph{True} indefinitely. For the second condition, we want our repeated application of $f$ to lead us into some fixed point, $y \in \texttt{fix}(f)$. Let us call the set of all such $y \in \texttt{fix}(f)$ for which the repeated application of $f$ on some initial value $x \in [A]_C$ converges to $y$, the set of \emph{Attractors} of the function $f$, denoted by $\texttt{fix}_A(x)$. We, then, notice that not all fixed points of $f$ are attractors and for all $z \in \texttt{fix}(f) \backslash \texttt{fix}_A(f)$, i.e. for the non-attractors of $f$, the only way to reach indefinite computation is the applicability of case 1 above. \\
	
	The only question that remains to be answered here is of knowing if the fixed points of a function exist. We have numerous fixed-point theorems that can provide help in this domain, and so this matter is not discussed here.
	
	\paragraph*{Existence of $k$-periodic points of $f$} 
	
	Similar to the discussion above, another way in which we can reach indefinite computation is when we start with some $x \in [A]_C$ such that the orbit of $x$ under the repeated application of $f$ is $k$-periodic and all $k$-values in this set satisfy the condition $C$. In other words, if for some $x\in [A]_C$, where $f^{(k)}(x)=x$ for some $k \geq 1$, we have $\{x, f(x), f(f(x)), \dots, f^{(k-1)}(x)\} \subseteq [A]_C$, then starting at $x$ will enter us into an infinite computation. The result is so trivial that a proof is not needed.
	 
	\paragraph*{Infinitely large domain and range} 
	
	In case the function $f$ has no fixed-points or $k$-periodic points, then the only way in which its repeated iteration will produce infinite computation is when we have $f^{(k)}(x) \in [A]_C$ for all $k \geq 0$. This will happen only when the range of $f$ is infinite, because no finite range one-one function can produce an infinite orbit with no repetitions of elements. The pigeon-hole principle won't allow for that to happen. Similarly, the domain of $f$ must also be infinite. However, the converse of this is not true, in general. \\
	
	Having studied the major reasons for encountering indefinite computation, an obvious question arises : Is there any computationally effective way to determine the set of all $k$-periodic points for all values of $k$? The answer is, no. However, not discouraging ourselves with this answer, it is still interesting to study properties of such infinitely truth preserving functions for one simple reason that infinite loops are way too common in routine programs and if we cannot identify them completely, we can at least study them to the maximum extent possible. For example, it is easy to see that condition weakening will have no effect on the order and limit of such functions, since we cannot go beyond infinity. Similarly, composition of two absolutely truth preserving functions must be an absolutely truth preserving function itself. A lot of such properties can be enumerated, but for the sake of brevity, the later sections in this report will expand more on this suitably. 
	
	\section{Maps between partial functions} 
	
	We have now studied enough theory about truth preservation functions to move a step closer to our actual aim of finding similarities between two sequential programs. So far, we have been limiting our discussion to functions of the kind $f:A \to A$, where the domain and range belonged to the same set. This was done primarily to study properties concerning iterative application of the function, for which it was necessary to have $Range(f) \subseteq Dom(f)$. Thus, the truth preservation of $f$ was limited to only one condition, $C$, that was imposed on the elements of $A$. We now discuss the case where the domain and range of $f$ are different sets. \\
	
	\begin{figure}[htp]
	\centering
	\begin{tikzpicture}
	\node (A) {$A$};
	\node (Ac) [below of=A, left of=A] {$[A]_{C_{1}}$};
	\node (B) [below of=Ac, right of=Ac] {$B$};
	\node (Bc) [node distance=5cm, right of=Ac] {$[B]_{C_{2}}$};
	
	\draw[->] (A) to node {$id_{C_{1}}$} (Ac);
	\draw[->] (Ac) to node {$f$} (B);
	\draw[->] (B) to node {$id_{C_{2}}$} (Bc);
	\draw[->] (A) to node {$f_{C_{1}}$} (B);
	\draw[->, dashed] (A) to node {$\Psi_{C_{1},C_{2}}$} (Bc);  
	\end{tikzpicture}
	\caption{Diagram to study maps between partial functions} 
	\label{fig:Fig2}
	\end{figure}
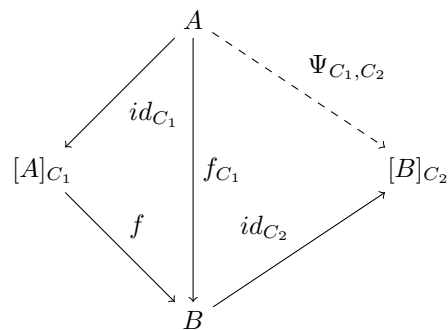
	
	Given two partial functions, we ask ourselves a question similar to what motivated us for Def. \ref{def:truth_Preserv}. Here, we focus only on a sub-diagram of Fig. \ref{fig:Fig1}, as given in Fig. \ref{fig:Fig2}. We defined $\Psi_{C_1,C_2}$ to be a partial function, defined only when $x \in [A]_{C_1}$ and $f(x) \in [B]_{C_2}$. The commutative nature of the diagram in Fig. \ref{fig:Fig2} allows us to write $\Psi_{C_1,C_2}=(id_{C_2} \circ (f \circ id_{C_1}))$. This way, all restrictions of $f$ can be removed and we need not concern ourselves with whether $f$ is total or partial. The two conditional identity functions take care of the undefinability of $f$ under appropriate situations. If we lift the two sets $A$ and $B$ to include the \emph{undefined} element, we only need to extend the definition of $f$ from $f:A \to B$ to $f:A_\perp \to B_\perp$ to obtain $\Psi_{C_1,C_2}:A_\perp \to B_\perp$ as the required partial function. In fact, lifting the sets converts all partial functions into total functions. Thus, from now on, all our analysis will assume that the domains of all partial functions have been lifted to convert them into total functions. \\
	
\begin{figure}[htp]
	\centering
	\begin{tikzpicture}
	\draw (-3,0) circle(2);
	\draw (3,0) circle(2);
	\draw (-3,-1) circle(0.75);
	\draw (3,1) circle(0.75);
	
	\node at (-3,1.5) (A) {$A_\perp$};
	\node at (3,-1.5) (B) {$B_\perp$};
	\node at (-3,-1) (Ac) {$[A_{\perp}]_{C_1}$};
	\node at (3,1) (Bc) {$[B_{\perp}]_{C_2}$};
	
	\draw[->, dashed] (A) to node {$id_{C_{1}}$} (Ac);
	\draw[->, dashed] (B) to node {$id_{C_{2}}$} (Bc);
	\draw[->, dashed] (Ac) to node {$f$} (B);
	\draw[->] (A) to node {$\Psi_{C_1,C_2}$} (Bc);
	\end{tikzpicture}
	\caption{Diagramatic representation of $\Psi_{C_1,C_2}$} 
	\label{fig:Fig3}
	\end{figure}
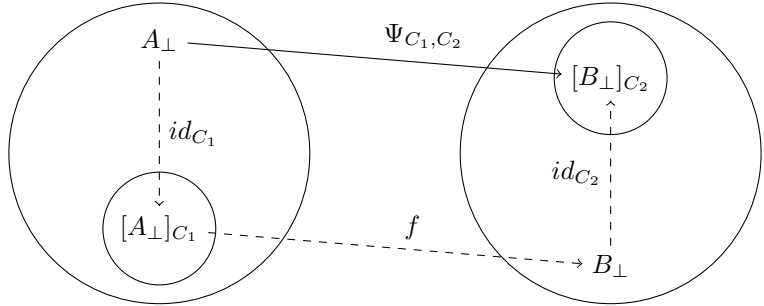

	The nature of $\Psi_{C_1,C_2}$ was explained by saying that the truth of $C_1$ was preserved by $C_2$, since only those elements in $A$ that satisfy $C_1$, were mapped to those elements in $B$ that satisfy $C_2$. Fig. \ref{fig:Fig3} illustrates this fact by representing the appropriate subset inclusion relations as well. 
	
	We now present a completely novel interpretation of what the diagram in Fig. \ref{fig:Fig3} represents. Since this report is concerned with dealing with sequential programs, it will make sense if we can somehow relate partial functions with computer programs. The later sections will provide all the details necessary to convert any sequential computer program into an equivalent partial function. For now, it suffices for us to assume that every program can indeed be uniquely represented by a partial function. Thus, from now on, we can use the terms partial function and computer program interchangeably, without any loss due to generalization. The lifting operation on the domain of partial functions is equivalent to forcing it to be defined on all elements of its domain, and so it can also be done for the case of computer programs. The state corresponding to $\perp$ will now depict a state of the machine, entering which no further meaningful task can be performed. The machine ceases to halt once it reaches this state and enters into a state of eternal computation. What it is trying to compute is of no importance at this stage. Thus, we can safely assume that our machine will halt in a finite time if anf only if some valid input, which will never land the machine into the undefined state, is provided and the output is produced by that input in finite time. It may be said that the machine can still run forever without reaching the undefined state, when it is running some absolutely truth preserving function. \\
	
	Back to our interpretation of Fig. \ref{fig:Fig3}, we can safely say that every sequential computer program, $P$, is defined on  only some inputs which lead the machine under consideration to halt in finite time. The set of all inputs is assumed to be of the form given in Eq. \ref{Eq:stateVarGen}.
	
	\begin{equation}
	\label{Eq:stateVarGen}
	\mathcal{S}_\perp = \{\perp\} \cup \{x_1:T_{x_1}, x_2:T_{x_2},\dotsc,\}
	\end{equation} 
	
	Here, $\perp$ is the special \emph{undefined} state, and $x_i$ is a variable of type $T_{x_i}$. The notion of types is well-defined in literature, so we will not go into defining it separately here. Thus, every program, $P$, is essentially a transformation of elements in $\mathcal{S}_\perp$. If we represent the subset, $S_P$, of $\mathcal{S}_\perp$ as the set of all valid inputs for $P$ (i.e. for which $P$ will make our machine halt in finite time), then we can represent the selection of this subset through specifying a condition, $C_P \in \mathcal{C}_{\mathcal{S}_\perp}$ such that $C_P = [ x \in S_P ]$. This way, $P$ is similar, in terms of the set of values it operates on, to $(g \circ id_{C_P})$, where $g$ is some function which transforms all $x \in [\mathcal{S}_\perp]_{C_P}$ to some set $\mathcal{S'}$ and the others to $\{\perp\} \in \mathcal{S'}_\perp$. Thus, the set of output state variables is $\mathcal{S'}_\perp$. Now, not all values in $\mathcal{S'}_\perp$ might be achievable by $P$, and this depends on how large $\mathcal{S'}_\perp$ is chosen. Only a subset of $\mathcal{S'}_\perp$, say $[\mathcal{S}_\perp]_{C'}$, is achievable by $P$, and we select this through the application of $id_{C'}$ on $\mathcal{S'}_\perp$. Thus, the whole program, $P$, given an input set of state variables, $\mathcal{S}_\perp$, essentially transforms only those variables that satisfy $C_P$ and produces values in a possibly different set of variables, $\mathcal{S'}_\perp$, conditioned on the fact that these values must satisfy $C'$. This way, $P$ acts as a truth preserving function from $\mathcal{S}_\perp$ to $\mathcal{S'}_\perp$, with respect to the conditions $C_P$ and $C'$, respectively.\\
	
	We can visualize $id_{C_P}$ and $id_{C'}$ as computer programs themselves, similar to that described in Fig. 
	\ref{Fig:idcCode}. In fact, any transformation of elements in $\mathcal{S}_\perp$ to those in $[\mathcal{S}_\perp]_{C_P}$ can be seen as an extension of $id_{C_P}$ in some sense. Thus, instead of seeing $\Psi_{C_P,C'}$ as a computer program, we can now view the transformations $P_1:\mathcal{S}_\perp \to [\mathcal{S}_\perp]_{C_P}$ and $P_2:\mathcal{S'}_\perp \to [\mathcal{S'}_\perp]_{C'}$ as our sequential programs, which may or may not halt for all inputs provided to them. The function $\Psi_{C_P,C'}$ performs a very critical operation now. The functions corresponding to $P_1$ and $P_2$, which for the sake of simple representation, are denoted by the same symbols, are related to each other through $\Psi_{C_P,C'}$. If we write $\Psi_{C_P,C'} = (id_{C'} \circ (P \circ id_{C_P}))$, or in this case, $\Psi_{C_P,C'} = (P_2 \circ (P \circ P_1))$, we go from the inputs of $P_1$ to the output of $P_2$ through an intermediate transformation of state variables, performed by $P$. In a way, there is a way to mimic the operations performed by $P_1$ through the operations performed by $P_2$ and $\Psi_{C_P,C'}$ proves this point for us. The situation is similar to sub-program isomorphism, in which $P_2$ is semantically isomorphic, i.e. performs the same transformation of its input variables, to $P_1$. The computation performed in $P_1$ is in some sense, similar to that performed in $P_2$ and once again, $\Psi_{C_P,C'}$ captures this similarity for us. Thus, the behavior of $P_1$ is replicated, in a stritly semantic sense, by $P_2$. This is the principle thought behind this thesis. \\
	
	With the above thought in mind, we can now write $\Psi_{C_P,C'}$ to be a map between $P_1$ and $P_2$ instead, and interpret it as establishing the required similarity in the behavior of the two programs. More specifically, $\Psi_{C_P,C'}$ alone would not suffice for this mapping. The transformation $P$ is equally important, since it is precisely this function which provides us with a way to relate the two possibly different state variable sets. Hence, we have to talk of $\Psi_{C_P,C'}$ and $P$ together, when trying to study similarity in computations. We denote this fact through the diagram in Fig. \ref{fig:Fig4}. In short, the relations (called \textit{arrows}) are precisely the pairs of the form $(P,\Psi_{C_P,C'})$, with some obvious generalizations, and the composition of arrows will be defined similar to function composition. The identity arrow will constitute $P = id_{\emph{True}}$ and $C_P = C'$.\\
	
	\begin{figure}[htp]
	\centering
	\begin{tikzpicture}
	\node (A) {$P_1:\mathcal{S}_\perp \to [\mathcal{S}_\perp]_{C_P}$};
	\node (B) [below of=A, node distance = 2cm] {$P_2:\mathcal{S'}_\perp \to [\mathcal{S'}_\perp]_{C'}$};
	
	\draw[->] (A) to node {$(P,\Psi_{C_P,C'})$} (B);
	\end{tikzpicture}
	\caption{Maps between functions} 
	\label{fig:Fig4}
	\end{figure}
	
\begin{figure}[htp]
	\centering
	\begin{tikzpicture}
	\node (A) {$P_1:\omega \to \{1,2,3,\dots,10\}$};
	\node (B) [below of=A, node distance = 2cm] {$P_2:\omega \to \{10,20,30,\dots,100\}$};
	
	\draw[->] (A) to node {$(110-10i,(P_2 \circ (P \circ P_1)))$} (B);
	\end{tikzpicture}
	\caption{Example for maps between functions} 
	\label{fig:Fig5}
	\end{figure}
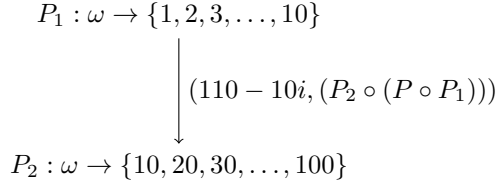

	Let us see an example of this formulation to better understand what is really going on here. We said that if there exists some pair of functions $(P,\Psi_{C_P,C'})$ between two given functions $P_1$ and $P_2$, then the nature of computation performed by these two functions, or equivalently computer programs, is similar. Assume $P_1$ to be a \texttt{while}-loop, which counts a variable $i \in I$, in some index set $I$, from $1$ upto $10$ in unit sized steps. Let $P_2$ be a \texttt{while}-loop, which counts down a variable $j \in J$, in some different index set $J$, from $100$ to $10$, in fixed steps of size $10$. Although the two programs perform seemingly different computations, the nature of this computation is similar : count up/down some index variable $10$ times. This similarity is captured through the function $P:I \to J$, by setting $j=P(i) = 110 - 10i$. If we set $\omega = \{0,1,2,3,\dots\}$, $C_1 = [1 \leq i \leq 10]$ and $C_2 = [j \in \{10,20,30,\dots,100\}]$, then $P_1$ maps elements from $\omega$ to those in $\omega_{C_1}$, and $P_2$ maps elements in $\omega$ to those in $\omega_{C_2}$. This is because we can interpret $P_1(i) = i+1$ for $i \in \{0,1,2,3,\dots,9\}$ and $P_1(i) = \perp$ for the remaining $i \in \omega$. A similar interpretation can be given for $P_2$. This way, $P$ establishes the required transformation from $I$ to $J$ and we can view the composition $(P_2 \circ (P \circ P_1))$ as capturing this notion of similar treatment with the index variables in the two programs. Diagramatically, we can represent this similar to Fig. \ref{fig:Fig4}, through Fig. \ref{fig:Fig5}.
	
\begin{figure}[htp]
	\centering
	\begin{tikzpicture}
	\node (A) {$P_1:\omega \to \{1,2,3,\dots,10\}$};
	\node (C) [below of=A, left of=A, node distance = 3cm] {$P_3:\omega \to \{1,2,4,8,\dots,512\}$};
	\node (B) [below of=C, right of=C, node distance = 2cm] {$P_2:\omega \to \{10,20,30,\dots,100\}$};
	
	\draw[->, bend left] (A) to node [right] {$(110-10i,(P_2 \circ (P \circ P_1)))$} (B);
	\draw[->, bend right] (A) to node [left] {$(2^i,(P_3 \circ (G \circ P_1)))$} (C);
	\end{tikzpicture}
	\caption{Adding another function to Fig. \ref{fig:Fig5}} 
	\label{fig:Fig6}
	\end{figure}
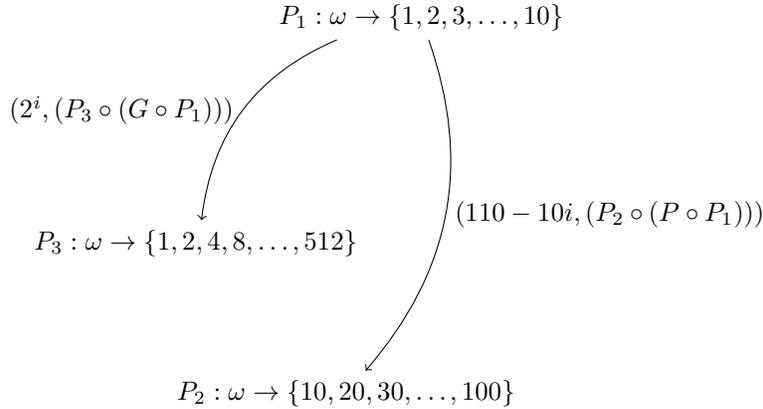

	Now, if we have a third program, $P_3$, which, say, counts a variable $k \in K$, in some other index set $K$, inside a \texttt{while}-loop from $1$ to $512$, by multiplying $k$ by $2$ everytime, it shouldn't be tough to see that the function $G:I \to K$, defined by $k=G(i)=2^i$, establishes the required map between $P_1$ and $P_3$. We can then represent this fact by adding another node corresponding to $P_3$ in our diagram above, to obtain the new diagram as in Fig. \ref{fig:Fig6}.
	
	The next obvious step is to, somehow, relate $P_3$ with $P_2$. This can be done either through mapping the elements in $\{10,20,\dots,100\}$ to those in $\{1,2,4,\dots,512\}$ directly, or going from the former to $\{1,2,3,\dots,10\}$ first, and then to $\{1,2,4,\dots,512\}$. We would prefer the latter approach, since this allows commutativity in the diagram above. However, for this to happen, we must find a way to go from $P_2$ to $P_1$ first. We can establish a map from $P_2$ to $P_1$ by noticing that the function $P$ is a bijection and thus, there exists an inverse, $P^{-1}$, which will do the job for us. Generalizing this, if the mapping between state variables is bijective for two programs, then so is the map which establishes the similarity between them. The diagram above, can now be made to commute by addition of a new arrow between $P_2$ and $P_3$, and reversing the one between $P_1$ and $P_2$, as shown in Fig. \ref{fig:Fig7}.\\
	
	\begin{figure}[htp]
	\centering
	\begin{tikzpicture}
	\node (A) {$P_1$};
	\node (C) [below of=A, left of=A, node distance = 3cm] {$P_3$};
	\node (B) [below of=A, right of=A, node distance = 3cm] {$P_2$};
	
	\draw[->, bend right] (B) to node [right] {$(\frac{110-j}{10},(P_1 \circ (P^{-1} \circ P_2)))$} (A);
	\draw[->, bend right] (A) to node [left] {$(2^i,(P_3 \circ (G \circ P_1)))$} (C);
	\draw[->, bend left] (B) to node [below] {$\left( 2^{(\frac{110-j}{10})},(P_3 \circ (H \circ P_2)) \right)$} (C);
	
	\end{tikzpicture}
	\caption{A commuting map between three programs} Note that $H = (G \circ P^{-1})$.
	\label{fig:Fig7}
	\end{figure}  
	
	As is evident from the diagram, reversing the arrow inverts the function $P$, as well as reverses the order of composition of functions. The order is $(P_1 \circ (P^{-1} \circ P_2))$ now, since we first perform the operations in $P_2$, and then perform the operations in $P_1$. Similarly, for the arrow between $P_2$ and $P_3$, note that the state variable transformation function, $H$, is nothing but a composition of the functions $G$ and $P^{-1}$. This ensures that the diagram in Fig. \ref{fig:Fig7} commutes. Also, since all three functions, $P, G$ and $H$ are bijections, the three arrows can be reversed appropriately. \\
	
	Before ending this section, an important clarification must be made regarding the notation of composition used above. A careful reader must have noticed the order of functions in $(P_2 \circ (P \circ P_1))$. This suggests that the transformation of the state variables happens after the program $P_1$ has completed its execution and the machine has come to a halt. As a consequence, mapping the index variable now will only change the last value taken by this index variable, instead of completely transforming all of them. This way, the program $P_2$ will only operate on at most one value and halt on the next step. However, this is not the case here. The composition operator is not really composing the functions represented by the two programs. In fact, we are never executing the two programs in any specified order. The notation just tells that we can reach from the input of $P_1$ to the output of $P_2$ if at every step that $P_1$ performs, we execute a corresponding step in $P_2$ by mapping the state variables of $P_1$ to those in $P_2$ using the transformation function $P$. The abused notation may confuse a some readers, in which case we can also use the notation $(P_2 \diamond (P \diamond P_1))$ to represent the same fact. \\
	
	\section{Arrows, in detail}
	
	The last section theorized a novel method to relate two partial functions, which perform similar operation upon the elements of their respective domains, through arrows, each of which consisted of a transformation function and a truth preserving function. This theory can directly be extended to sequential computer programs to study similarity in the nature of computation performed by a given set of such programs. However, before moving further with this extension of concepts, we get into a few more details about the nature of arrows connecting two partial functions. Two more types of arrows will be defined to capture the notions of sub-structure isomorphism within programs and Turing reducibility of one formal language into another, with special cases of mapping reducibility and polynomial time reducibility. With this, the diagrammatic representations containing these arrows will capture the actual sense of similarity between two given programs. \\
	
	Let us first introduce some abstractions and notations that we intend to follow throughout the discussion following this point. Basically, two models of computation, the first one being a specific form of the second, are discussed. The final results are derived for both the models.
	
	\paragraph*{Model 1 :}The first model of computation is close to a real sequential machine, consisting of four basic modules: data storage, instruction storage, control unit and a set of I/O ports. The control unit consists some local memory, in which the instructions of a program, stored in the intruction storage, are fetched one after the other, in order of their appearance in the program. This local memory can also store some data, which is required to execute the current instruction, as pointed by a program counter. The data can come from either the data storage or from the users through any of the I/O ports. The instructions of a given program cannot be altered during the program's execution. Thus, when the control unit executes some instruction, the only entities that change are the data bits in the local memory of this unit or of the main data storage. The set of all these bits, which can be changed during a program's execution, when assigned a particular value to each bit, is called a \emph{state assignment} of the machine.\\
	
	The control unit is believed to work using some function, or more precisely, a finite state automaton, built into the hardware or known to the unit in some way, which may change the state assignment of the machine to a different one, in which the mutable bits have been assigned different values. The set of all possible values these bits can take is called the \emph{state variable set}. Thus, every instruction in a given program is mapping from this state variable set to itself. The language in which the instruction is written is unimportant as long as the control unit is able to parse it for us. We can give one particular assignment (say, an assignment of all zeros or all ones : not important) a special name, say \emph{undefined}, which is interpreted as an invalid state for the machine. By this, we mean that if the machine ever enters this state assignment, it will never be able to come out of it and cease to halt. We say that our machine halts after executing a program if it completes the execution of its last instruction ending in some valid state assignment.\\
	
	With this model in mind, it may seem that the set of state variables is different from the one we defined in Eq. \ref{Eq:stateVarGen}. However, it is not. The notion of types in Eq. \ref{Eq:stateVarGen} are nothing but groupings of the mutable bits we talked of, based on some abstract notion. The state, $\perp$, is the same as the invalid assignment that we talked of in the above paragraph. Thus, we remain consistent with our computation model throughout. Denote a given computer program by $P$ and the set of its state variables by $\mathcal{S}_\perp$. This way, we can visualize our program as $P:\mathcal{S}_\perp \to \mathcal{S}_\perp$. Each instruction in a given program is a program in itself, and since all instructions perform some transformation of the states, $P$ can be interpreted like a function as well. The two terms, function and program, will hence, be used interchangeably. \\
	
	The set of valid state assignments for which our program $P$ halts in a valid state in finite time is denoted by $\mathcal{S}_\perp'$. Clearly, we have $\mathcal{S}_\perp' \subseteq \mathcal{S}_\perp$, and hence, we can write an equivalent Boolean condition $C \in \mathcal{C}_{\mathcal{S}_\perp}$, to select this subset our of our set of state variables. This will allow us to write $\mathcal{S}_\perp' = [\mathcal{S}_\perp]_C$, consistent with the notations used so far. The function $P$ is total if $C = \emph{True}$ and partial otherwise. In case $C$ is not \emph{True}, the partial definition of $P$ can be converted into total definition, i.e. $P$ can be converted into a total function, by setting $P(x)=\perp$ for all $x \in \mathcal{S}_\perp \backslash [\mathcal{S}_\perp]_C$. \\
	
	Given a set of programs, index the elements of this set as $\{P_1, P_2, P_3, \dots \}$. The domains and conditions for these programs will be indexed similarly. Thus, the domain of $P_1$ is $\mathcal{S}_{1,\perp_1}$. We index the special state assignment, $\perp$, as well, to stress on the fact that different state variable sets can be lifted by different state assignments and hence, the bottom elements in all these sets may not be the same. For any two given programs, or functions, say $P_i$ and $P_j$, denote the transformation function between its state variables by $T_{i,j}: \mathcal{S}_{i,\perp_i} \to \mathcal{S}_{j,\perp_j}$. Unless otherwise specified, assume that this function always exists. If the set $\mathcal{S}_{j,\perp_j}$ has larger cardinality than $\mathcal{S}_{i,\perp_i}$, then assume $T_{i,j}$ to be injective, unless otherwise specified. For reverse other case, assume that $T_{i,j}$ is surjective, and for the case where the two cardinalities are same, assume that $T_{i,j}$ is bijective, unless otherwise specified.\\
	
	We also assume that the state variable set is totally ordered, i.e. there exists a relation $\prec_{\mathcal{S}_{i,\perp_i}}$ such that for all $x_i, y_i \in \mathcal{S}_{i,\perp_i}$, either $x_i \prec_{\mathcal{S}_{i,\perp_i}} y_i$ or $y_i \prec_{\mathcal{S}_{i,\perp_i}} x_i$ or $x_i = y_i$. Accordingly, assume that the program $P_i$ induces a total order on these elements, say a \emph{Program order}, denoted by $\prec_{P_i}$, such that for all $x_i, y_i \in \mathcal{S}_{i,\perp_i}$, we have $x_i \prec_{S_{i,\perp_i}} y_i$ implies $P_i(x_i) \preceq_{P_i} P_i(y_i)$, i.e. the program $P_i$, when seen as a mapping $P_i:(\mathcal{S}_{i,\perp_i},\preceq_{\mathcal{S}_{i,\perp_i}}) \to (\mathcal{S}_{i,\perp_i},\preceq_{P_i})$, is order preserving (may not be strictly monotonic). However, this may not mean that $P_i$ is monotonic with respect to $\prec_{\mathcal{S}_{i,\perp_i}}$, i.e. for $x_i \prec_{\mathcal{S}_{i,\perp_i}} y_i$, we may not have $P_i(x_i) \prec_{\mathcal{S}_{i,\perp_i}} P_i(y_i)$. The order induced by $P_i$ on the elements of $\mathcal{S}_{i,\perp_i}$ may be different from the inherent order, $\prec_{\mathcal{S}_{i,\perp_i}}$, amongst the elements of this set. 
	
	\paragraph*{Model 2 (Abstraction of Model 1) :}For a more formal discussion, as will be required for type-{2} arrows in particular, we extend the scope of functions discussed so far to the class of computable functions, replacing the term \emph{program} with the more technical term, \emph{Turing Machine}, written as $TM$ in short. Within this model, we now talk of recognizability of formal languages defined on the alphabet, $\Sigma = \{0,1\}$, where each language, $L$, is some subset of $\Sigma^\ast$, the Kleene closure of $\Sigma$, representing the set of all strings that can be formed out of the letters in $\Sigma$. The notion of functions is then interpreted as characterizing these languages so that the relation between functions, as defined by the different types of arrows, can be extended to relations between different languages.\\
	
	The set $\Sigma^\ast$ can also be interpreted as the domain of all functions considered so far, since all state variable assignments can be uniquely encoded using only $0$s and $1$s, and hence, form elements of $\Sigma^\ast$. Similarly, the range of each of these functions is a subset of $\Sigma^\ast$ as well. Thus, if for $P_i : \mathcal{S}_{i,\perp_i}$, we define the corresponding language, $L_i$ to be $\{\omega \mid \omega \in Im(P_i)\}$, then the Turing-recognizability of $L_i$ implies the  computability of $P_i$. Now, since there can be many Turing machines that enumerate the strings in $L_i$, i.e. recognize $L_i$, each one of them is similar to a computer program to compute $P_i$. We can also represent all conditions defined on state variable assignments in terms of properties of strings in $\Sigma^\ast$. To say that a set of state variable assignments satisfy some given condition, $C$, is equivalent to saying that the strings in $\Sigma^\ast$, that correspond to the encodings of these assignments, have the property of satisfying the condition, $C$. Assuming that checking $C$ on any assignment, or equivalently, checking if a given string $\omega \in \Sigma^\ast$ follows property, $P$, is performed through some checking function, $P_C$, which returns \emph{True} if $C(\omega)$ is \emph{True}, and \emph{False} otherwise, the conditional identity function, $id_C$, becomes computable if $P_C$ is computable. This can be proved by noticing that the set, $\Sigma^\ast$ is recursively enumerable, and thus, an easy method to compute $id_C$ is to take every string in $\Sigma^\ast$ one by one, run $P_C$ on it and return its output. This way, $id_C$ is polynomial time reducible to $P_C$, and accordingly, every $f_C = (f \circ id_C)$, for some $f$, is computable if and only if both $f$ and $id_C$ are computable.\\
	
	We are now ready to study the different kinds of arrows that can exist between two total functions defined on lifted domains, or equivalently, two Turing machines with given specifications. The results derived for type-$0$ and type-$1$ arrows in model $1$ will be suitably extended to Model $2$ and type-$2$ arrows will then be formulated.
	
	\subsection{Type-$0$ arrow : Transformational isomorphism}
	
	Let us start with the kind of arrows that we saw in the last section. We call them having type-$0$ because they are the trivial-most kind with respect to the extensions that we will see later. Although the concept behind this kind of arrows has been introduced previously, we intend to give formal definitions in this section, for the sake of removing all ambiguity that may arise later.\\
	
	The motivation behind type-{0} arrows is the representability of every program inside an appropriately designed \texttt{while}-loop, using some auxillary variables, if necessary, and then studying the termination properties of that program with respect to the termination of the loop that envelopes it. Consider the general \texttt{while}-loop as given in Fig. \ref{Fig:loop}. If the domain from which $x$ is selected is extended to the state variable set, $\mathcal{S}_\perp$, then an equivalent computable function (as long as $f$ is computable and checking the condition $C$ on any state variable assignment is possible in finite time), $\mathcal{L}^{f}_{C}:\mathcal{S}_\perp \to \mathcal{S}_\perp$ can be given as:
	
	\begin{equation}
	\label{Eq: loopFormal}
	\mathcal{L}^f_C(x) =
	\begin{cases}
	\mathcal{L}^f_C(f(x)) & \text{ if }C(x)\text{ is \emph{True}}\\
	x & \text{ otherwise}
	\end{cases} 
	\end{equation}
	
	The termination of this recurrence is entirely dependent on the truth preservation order of $f$ with respect to the condition $C$. Thus, we can always choose $x$ and $C$ in such a way that the loop runs for exactly for some given number of times, and this way, we can surround any program with a \texttt{while}-loop. For example, if some statement in any program gets executed exactly once, extend the state variable set to include an auxillary state variable, which during the initialization for the loop, gets the value $0$. Let the condition $C$ check when this variable takes a value other than $0$ and extend the definition of $f$ to one higher dimension, in which the value if this variable is incremented by one. This way, if the other state variables are initialized appropriately so that the statement does execute, then the application of $f$ will cause our auxillary variable to take the value $1$ and hence, terminate the loop immediately. Thus, we have ensured that the function $f$ gets applied exactly once.\\
	
	The prime reason for writing all programs inside suitably designed \texttt{while}-loops is to study their termination properties using the number of times the instructions in that program get executed. The aim is not exactly to know when a program would stop execution, but to know if two given programs, when started at the same time on some inputs, would execute same number of instructions before halting, and then using the known termination properties of one program to deduce facts about the other. The example taken in the last section of the previous section highlights this fact more closely. The three loops inside $P_1$, $P_2$ and $P_3$ carry the same number of operations for all initializations of the index variables and hence, are similar in this aspect of the extent of computation performed on any input. Type-$0$ arrows try to capture this notion, more formally.\\
	
	\begin{mydef}
	\label{def:type0arrow}
	Let $C_i \in \mathcal{C}_{\mathcal{S}_{i,\perp_i}}$ and $C_j \in \mathcal{C}_{\mathcal{S}_{j,\perp_j}}$ be two Boolean conditions. For two given programs $P_i:\mathcal{S}_{i,\perp_i} \to \mathcal{S}_{i,\perp_i}$ and $P_j: \mathcal{S}_{j,\perp_j} \to \mathcal{S}_{j,\perp_j}$, let an injective (one-one) order-preserving transformation function between their ordered domains be $T_{i,j}:\left( \mathcal{S}_{i,\perp_i}, \preceq_{S_{i,\perp_i}} \right) \to \left( \mathcal{S}_{j,\perp_j}, \preceq_{S_{j,\perp_j}} \right)$. Then there exists an \textbf{arrow of type-$0$}, $A^0_{i,j} \in \mathcal{A}^0_{i,j}$, from $\mathcal{L}^{P_i}_{C_i}$ to $\mathcal{L}^{P_j}_{C_j}$, denoted by $A^{0}_{i,j}: \langle T_{i,j} \rangle$, if the truth preservation orders of all elements in the domains of $P_i$ and $P_j$ are the same, with respect to $C_i$ and $C_j$, respectively, i.e. 
	\begin{equation}
	\label{Eq:type0def}
	\Theta_{C_i,P_i}(x) = \Theta_{C_j,P_j}(T_{i,j}(x)) \quad \forall x \in S_{i,\perp_i}
	\end{equation}
	
	Here, $\mathcal{A}^0_{i,j}$ is the set of all arrows of type-0 that can exist between $\mathcal{L}^{P_i}_{C_i}$ and $\mathcal{L}^{P_j}_{C_j}$, one for each order-preserving injection, $T_{i,j}$. The existence of $A^0_{i,j}$ establishes a \textbf{Transformational Isomorphism} between $\mathcal{L}^{P_i}_{C_i}$ and $\mathcal{L}^{P_j}_{C_j}$, denoted by $\mathcal{L}^{P_i}_{C_i} \overset{0}{\sim} \mathcal{L}^{P_j}_{C_j}$.\\
	\end{mydef}  
	
	\begin{figure}[htp]
	\centering
	\begin{tikzpicture}
	\node (A) {$\mathcal{L}^{P_i}_{C_i}:\mathcal{S}_{i,\perp_i} \to \mathcal{S}_{i,\perp_i}$};
	\node (B) [below of=A, node distance = 2cm] {$\mathcal{L}^{P_j}_{C_j}: \mathcal{S}_{j,\perp_j} \to \mathcal{S}_{j,\perp_j}$};
	
	\draw[->] (A) to node {$A^0_{i,j}:\langle T_{i,j} \rangle$} (B);
	\end{tikzpicture}
	\caption{Diagrammatic representation of type-$0$ arrow} 
	\label{fig:type0arrowDef}
	\end{figure}
	
	The definition above is consistent with the formulation in the previous section. Diagrammatically, the arrow just described is denoted as $\mathcal{L}^{P_i}_{C_i} \xrightarrow{A^0_{i,j}} \mathcal{L}^{P_j}_{C_j}$, or with more details as in Fig. \ref{fig:type0arrowDef}. The \textit{inverse arrow} of $A^0_{i,j}$, given by $A^0_{j,i}: \langle T^{-1}_{i,j} \rangle $, exists and is of type-$0$, if $T_{i,j}$ is a bijection, i.e. it is invertible. \\
	
	The conditions specified in the definition seem hard to visualize at first, but are quite easy to understand. An injective map gaurantees that for every state assignment in $\mathcal{S}_{i,\perp_i}$, we have exactly one state assignment in $\mathcal{S}_{j,\perp_j}$. This is required to ensure a unique mapping of each state assignment, so that the nature of computation can be verified to be similar at every step of computation. However, no restriction has been placed on the relation in cardinalities of $\mathcal{S}_{i,\perp_i}$ and $\mathcal{S}_{j,\perp_j}$, due to which $T_{i,j}$ is not required to a bijection. We require the truth preservation of $C_i$ and $C_j$ by the transformation function $T_{i,j}$, as is evident from the nature of this function, to ensure that we do not tranform any assignment from $\mathcal{S}_{i,\perp_i}$ to some assignment in $\mathcal{S}_{j,\perp_j}$, which is considered invalid with respect to the application of $P_j$. This condition also ensures that the elements in $[\mathcal{S}_{i,\perp_i}]_{C_i}$ map to only those in $[\mathcal{S}_{j,\perp_j}]_{C_j}$. This way, the set of outputs of the two programs, at every step of computation, differ only in the names of their state variables.\\
	
	The most important condition is given by Eq. \ref{Eq:type0def}, which really establishes the similarity in the nature of computation performed by $P_i$ and $P_j$. This condition talks of a requirement of invariance of the truth preservation order under the transformation function. Combined with the other properties of $T_{i,j}$, this added restriction assures that for every computational step that $P_i$ performs, we can write a unique step performed by $P_j$. For any $x$, this condition means that the number of times $\mathcal{L}^{P_i}_{C_i}$ executes on $x$ is exactly same as the number of times $\mathcal{L}^{P_j}_{C_j}$ executes on $T_{i,j}(x)$. For example, counting from $1$ to $10$ in unit steps and counting from $1$ to $512$ by multiplying by $2$ repeatedly has a similar nature of computation. Arrows of type-$0$ capture precisely this kind of similarity, and hence, are said to establish a transformational isomorphism. In case such an isomorphism exists, we say that $\mathcal{L}^{P_i}_{C_i}$ is tranformationally isomorphic to, or simply, isomorphic to $\mathcal{L}^{P_j}_{C_j}$. Seen graphically, the data flow as well as the control flow in two programs is the same, and so is the number of operations performed in them. \\
	
	We saw how to compose two arrows in the discussion around Fig. \ref{fig:Fig7}. Formally, given two arrows $A^0_{i,j}$ and $A^0_{j,k}$, we define the composite arrow, $A^0_{i,k}$, as establishing transformational isomorphism between $\mathcal{L}^{P_i}_{C_i}$ and $\mathcal{L}^{P_k}_{C_k}$. For this to exist, we require the composite transformation function, $(T_{j,k} \circ T_{i,j})$ to satisfy all conditions as specified in Def. \ref{def:type0arrow}. Since, $T_{i,j}$ and $T_{j,k}$ are one-one functions, so will be $T_{i,k}$. Thus, the condition of injectivity is satisfied. In general, if $f:A \to B$ is a truth preserving function with respect to $C_A$ and $C_B$, and $g:B \to C$ is also a truth preserving function with respect to $C_B$ and $C_C$, then it can be easily shown that the composition $(g \circ f)$ is truth preserving with respect to $C_A$ and $C_C$. Hence, the arrow $A^0_{i,k}:\langle T_{i,k}\rangle$ must exist. We can also represent this fact as follows:
	
	\begin{equation}
	\label{eq:type0arrowtrans}
	 \left( \mathcal{L}^{P_i}_{C_i} \overset{0}{\sim} \mathcal{L}^{P_j}_{C_j} \right) \land \left( \mathcal{L}^{P_j}_{C_j} \overset{0}{\sim} \mathcal{L}^{P_k}_{C_k} \right) \implies \left( \mathcal{L}^{P_i}_{C_i} \overset{0}{\sim} \mathcal{L}^{P_k}_{C_k} \right)
	\end{equation}
	
	The equation above is true for all $\mathcal{L}^{P_i}_{C_i}, \mathcal{L}^{P_j}_{C_j}, \mathcal{L}^{P_k}_{C_k} \in \mathcal{P}$. The arrow $A^0_{i,i}:\langle id_{\emph{True}} \rangle $ is the \textit{identity arrow} for the program $\mathcal{L}^{P_i}_{C_i}$. The existence of this arrow proves that $\overset{0}{\sim}$ is reflexive. We already saw its transitive nature through Eq. \ref{eq:type0arrowtrans}. To prove its symmetric nature, we note that condition of injectivity in Def. \ref{def:type0arrow} implies that if we change the range of $T_{i,j}$ to $\mathcal{S}_{j,\perp_j}'$, where $[\mathcal{S}_{j,\perp_j}]_{C_j} \subseteq \mathcal{S}_{j,\perp_j}' \subseteq \mathcal{S}_{j,\perp_j}$ such that $T_{i,j}$ is now a bijection, an inverse arrow will exist from $\mathcal{L}^{P_i}_{C_i}$ to $\mathcal{L}^{P_j}_{C_j}$. This is allowed as long as a smaller range does not affect any transformations of valid states. If the only ones left out are some invalid states that are no longer mapped by $T_{i,j}$, then an inverse arrow is apparent. Since this does not affect the similarity in the nature of computation performed by $\mathcal{L}^{P_i}_{C_i}$ and $\mathcal{L}^{P_j}_{C_j}$, the transformational isomorphism is symmetric in nature, an hence, an equivalence relation on $\mathcal{P}$ in this case. However, in general, only transitivity and reflexivity is shown by transformational isomorphism.\\
	
	Stritly speaking, the restriction of a program's range, as done in the previous paragraph, is potential of causing minor inconvinience to an observant reader. We can be sure that reducing the domain will preserve the set of inputs of $\mathcal{L}^{P_j}_{C_j}$ for which the machine on which $\mathcal{L}^{P_j}_{C_j}$ runs will halt in finite time, it does not preserve the set of inputs for which the machine never halts. There may be some more state assignments for which the machine never halts on running $\mathcal{L}^{P_j}_{C_j}$, for which no corresponding state assignments in $\mathcal{S}_{i,\perp_i}$ exist, as is expected of the notion of isomorphism. However, if we restrict our analysis to the set of inputs for which the machine will halt, then the symmetric nature of $\overset{0}{\sim}$ is apparent.\\
	
	The transformation function in Def. \ref{def:type0arrow} has many more properties. Two of these are of particular interest to us. For all elements in $[S_{i,\perp_i}]_{C_i}$, there is a unique element in $[S_{j,\perp_j}]_{C_j}$ due to injectivity of $T_{i,j}$. Hence, every element in $S_{i,\perp_i}$, which has a non-negative truth preservation order with respect to $C_i$ also has a non-negative truth preservation order with respect to $C_j$, under the transformation. This makes $(id_{C_j} \circ (T_{i,j} \circ id_{C_i}))$ a truth preservating function with respect to $C_i$ and $C_j$. The injectivity of $T_{i,j}$ also makes it an order preserving mapping for the program orders of $\mathcal{L}^{P_i}_{C_i}$ and $\mathcal{L}^{P_i}_{C_i}$. The isomorphism, so established, is independent of the programming language used to code the programs. This is because we only concern ourselves with the input to output transformation a program simulates, for which the encoding of this transformation, or even the inputs and the outputs is unimportant, as long as we maintain coherence in our representation, i.e. we encode our inputs and outputs in such a way that the transformation function is suitably defined. We are also working in an almost hardware-independent space here, since only the nature of computation is compared. The example taken before relates addition with multiplication, and then with exponentiation. We can perform similar operations with modulo, division, subtraction or for that sake, any binary relation defined appropriately on the domain set, as long as all of these can be expressed as single instructions. Thus, the question of "what" is computed is less important than "how" it is computed, which is why, essentially, the latter is captured by transformational isomorphism.\\
	
	The existence of type-$0$ arrow between $\mathcal{L}^{P_i}_{C_i}$ and $\mathcal{L}^{P_j}_{C_j}$ is a stronger relation than the mapping reducibility of $L_i$ and $L_j$, the languages recognized by $\mathcal{L}^{P_i}_{C_i}$ and $\mathcal{L}^{P_j}_{C_j}$, respectively. In other words, if $\mathcal{L}^{P_i}_{C_i}$ and $\mathcal{L}^{P_j}_{C_j}$ are transformationally isomorphic to each other, then $L_i$ is mapping-reducible to $L_j$, but the reverse may not always be true. The simplest argument, perhaps, to show this is the restriction of same number of intructions, or computational steps, to establish transformational isomorphism, whereas, no such restriction exists for mapping-reducibility. As long as for every $\omega$, we have $\omega \in L_i \xLeftrightarrow[]{} f(\omega) \in L_j$, for some computable function, $f$, we say that $L_i \leq_m L_j$, i.e. $L_i$ is mapping-reducible to $L_j$. Since there can be many Turing machines that recognize a language, there is always a way to choose two machines, one each for $L_i$ and $L_j$, so that the number of steps performed by them is different. Hence, even though $L_i \leq_m L_j$, the Turing machines, thus chosen, or equivalently, the programs $\mathcal{L}^{P_i}_{C_i}$ and $\mathcal{L}^{P_j}_{C_j}$ are not transformationally isomorphic in this case.\\
	
	In fact, transformational isomorphism is quite a strong relation between two programs. For a given task, two versions of the same program to carry out that task may not be transformationally isomorphic. Even worse, two seemingly unrelated programs to carry out the same task are, in general, not transformationally isomorphic to each other. For example, any array of finitely many elements can be sorted using either Bubble sort or Merge sort, both of which perform different type of operations to output similar results. However, a difference in the worst case time complexity of the two sorting algorithms shows that the number of steps taken by them to halt is not the same for the outputs they produce. Hence, the programs to simulate Bubble sort and Merge sort  cannot be transformationally isomorphic to each other. With all this discussion, it seems very unlikely that there can ever exist any formal method, or equivalently a Turing machine, which can check if two given programs are isomorphic to each or not. The theorem \ref{thm:type0nonRE} shows that this really is the case, and indeed, no such Turing machine exists.\\
	
	Let $M_i$ and $M_j$ be the encodings, in $\Sigma^\ast$, of the our programs $\mathcal{L}^{P_i}_{C_i}$ and $\mathcal{L}^{P_j}_{C_j}$, respectively. The languages accepted by these programs, or equivalently, Turing machines, are $L(\mathcal{L}^{P_i}_{C_i})$ and $L(\mathcal{L}^{P_j}_{C_j})$, respectively. Let the number of instructions, as defined by the truth preservation order of any element, in $\mathcal{L}^{P_i}_{C_i}$ be given by $\mid \mathcal{L}^{P_i}_{C_i} \mid$, similarly for $\mathcal{L}^{P_j}_{C_j}$. Define two languages, $EQ_{TM}$ and $Ar^0_{TM}$, as follows:
	\begin{equation}
	\label{eq:eqtm}
	EQ_{TM} = \{ <M_i,M_j> \mid L(\mathcal{L}^{P_i}_{C_i}) = L(\mathcal{L}^{P_j}_{C_j}) \}
	\end{equation}
	\begin{equation}
	\label{eq:a0tm}
	Ar^0_{TM} = \{ <M_i,M_j> \mid \mathcal{L}^{P_i}_{C_i} \overset{0}{\sim} \mathcal{L}^{P_j}_{C_j} \}
	\end{equation}
	
	The definition of type-$0$ arrows can be used to expand upon Eq. \ref{eq:a0tm} as:
	
	\begin{equation}
	\label{eq:a0tm_2}
	Ar^0_{TM} = \{ <M_i,M_j> \mid L(\mathcal{L}^{P_i}_{C_i}) \leq_m L(\mathcal{L}^{P_j}_{C_j}) \text{ and } \mid \mathcal{L}^{P_i}_{C_i} \mid = \mid \mathcal{L}^{P_j}_{C_j} \mid \}
	\end{equation}
	
	We use the abused notation for the number of operations in $\mathcal{L}^{P_i}_{C_i}$ and $\mathcal{L}^{P_j}_{C_j}$ to denote that the truth preservation orders for each element in their respective domains is invariant under the transformation function. In the realm of Turing machines, this translates to a one-one correspondance between the number of state transitions taken by $M_i$ and $M_j$ for any string in $L(\mathcal{L}^{P_i}_{C_i})$ and its equivalent in $L(\mathcal{L}^{P_j}_{C_j})$. Thus, strictly speaking, the equality above is not really a strict equality, but rather a weak similarity in the number of state transitions each Turing machine performs.\\
	
	\begin{mythm}
	\label{thm:type0nonRE}
	The language $Ar^0_{TM}$ is non-Turing-recognizable, i.e. it is not recursively enumerable.
	\end{mythm} 
	
	\paragraph*{Proof} To show the non-recursive enumerability of $Ar^0_{TM}$, the best strategy, perhaps, would be reduce some non-recursively enumerable language into it. This way, the fact that no Turing machine exists for the language reduced to $Ar^0_{TM}$ would help us deduce the same for our language as well. For the proof here, we try to reduce $EQ_{TM}$ to $Ar^0_{TM}$, since we know that the former is non-Turing recognizable. Assume that the Turing machines under consideration consist of a work tape along with a print tape, similar to an enumerator machine, so that all the output appears on the print tape and the computations are performed on the work tape. Now, our aim is to use a solver for $Ar^0_{TM}$ to solve $EQ_{TM}$, which means that we must assume that some Turing machine, $M$, indeed, accepts $Ar^0_{TM}$. If we have to check for two given languages, $L(M_1)$ and $L(M_2)$, for their equality, we first observe that $L(M_1) = L(M_2)$ implies $L(M_1) \leq_m L(M_2)$. Thus, when converted into an instance of $M$, the first condition of mapping reducibility will automatically be satisfied. The second condition, however, may not always be true. Two Turing machines accepting the same language need not do that in the same number of state transitions. However, since we know that the machines do accept the strings in the languages, they must do this in finite number of transitions. Hence, if during the acceptance of a string, say $\omega$, the machine $M_1$ makes $T_1(
	\omega)$ transitions and $M_2$ makes $T_2(\omega)$ transitions, we can modify the control units for these machines in such a way that as soon as one of them is ready to print $\omega$ on the output tape, it keeps on looping in the same state until the other machine also becomes ready to print $\omega$ on its output tape. When both the machines are ready with their computation, they print $\omega$ at the same time. The modified nature does not disturb the languages accepted by the machines and preserves the mapping reducibility relation because $\mid T_1(\omega) - T_2(\omega) \mid$ is computable. This way, we have converted the instance of $EQ_{TM}$ into an instance of $Ar^0_{TM}$, and the modified machines are, actually, transformationally isomorphic now. Hence, the non-Turing recognizability of $EQ_{TM}$ is carried over to $Ar^0_{TM}$ and we conclude the proof here.\\
	
	In the next subsection, we take this notion of isomorphism between two programs to the next level, where we, then, call $\mathcal{L}^{P_i}_{C_j}$ and $\mathcal{L}^{P_j}_{C_j}$ similar if $\mathcal{L}^{P_i}_{C_i}$ is transformationally isomorphic to some code inside $\mathcal{L}^{P_j}_{C_j}$. Let us see how this arrow is formulated.
	
	\subsection{Type-$1$ arrow : Sub-structure transformational isomorphism}
	
	The arrows we just saw establish isomorphisms between programs performing similar computations, on possibly different state variable sets. This is helpful as long as for every step that $\mathcal{L}^{P_1}_{C_1}$ performs, we are able to write a corresponding step that $\mathcal{L}^{P_2}_{C_2}$ performs. The number of steps is the same for two programs (only for the inputs belonging to the domain of $\mathcal{L}^{P_1}_{C_1}$. The second program, $\mathcal{L}^{P_2}_{C_2}$ may have many more valid input state assignments on which it can take any number of steps for execution). However, this may not be the case always. In fact, in most cases, there will only be a small part of $\mathcal{L}^{P_2}_{C_2}$ that is similar to the computation performed by $\mathcal{L}^{P_1}_{C_1}$, and not the whole program itself. For example, if $\mathcal{L}^{P_1}_{C_1}$ is a loop, counting down from $10$ to $1$ in unit steps, and $\mathcal{L}^{P_2}_{C_2}$ is a computer program which contains a loop counting up from $50$ to $100$ in unit steps, surrounded by some other computation (like function calls etc.) then $\mathcal{L}^{P_2}_{C_2}$ contains a loop that counts up in ten unit steps, surroundedm obviously, by the code to make up for the remaining computation.  Thus, only a part of $\mathcal{L}^{P_2}_{C_2}$ is transformationally isomorphic to $\mathcal{L}^{P_1}_{C_1}$. This is exactly the notion captured by arrows of type-$1$, and hence, they are said to establish a \emph{Sub-Structure Transformational Isomorphism} between two programs. In the definition below, assume that $(\mathcal{L}^{P_i}_{C_i} \circ \mathcal{L}^{P_j}_{C_j})$ means the program $\mathcal{L}^{P_j}_{C_j}$ is followed by the program $\mathcal{L}^{P_i}_{C_i}$, i.e. the output of $\mathcal{L}^{P_j}_{C_j}$ is fed into the input of $\mathcal{L}^{P_i}_{C_i}$. This can be written as a concatenation of the code for $\mathcal{L}^{P_j}_{C_j}$ and $\mathcal{L}^{P_i}_{C_i}$, denoted as $\mathcal{L}^{P_j}_{C_j};\mathcal{L}^{P_i}_{C_i}$, or $\mathcal{L}^{P_i}_{C_i}(\mathcal{L}^{P_j}_{C_j})$.\\
	
	\begin{mydef}
	\label{def:type1arrow}
	Let $C_i \in \mathcal{C}_{\mathcal{S}_{i,\perp_i}}$, $C_j \in \mathcal{C}_{\mathcal{S}_{j,\perp_j}}$, $C_k \in \mathcal{C}_{\mathcal{S}_{k,\perp_k}}$ and $C_l \in \mathcal{C}_{\mathcal{S}_{l,\perp_l}}$ be four Boolean conditions. For two given programs $\mathcal{L}^{P_i}_{C_i}:\mathcal{S}_{i,\perp_i} \to \mathcal{S}_{i,\perp_i}$ and $\mathcal{L}^{P_j}_{C_j} = (\mathcal{L}^{P_m}_{C_m} \circ (\mathcal{L}^{P_k}_{C_k} \circ \mathcal{L}^{P_l}_{C_l}))$, for some $\mathcal{L}^{P_k}_{C_k}, \mathcal{L}^{P_l}_{C_l}, \mathcal{L}^{P_m}_{C_m} \in \mathcal{P}$ and $\mathcal{L}^{P_l}_{C_l}:\mathcal{S}_{l,\perp_l} \to \mathcal{S}_{l,\perp_l}$, then there exists an \textbf{arrow of type-$1$}, $A^1_{i,j} \in \mathcal{A}^1_{i,j}$, from $\mathcal{L}^{P_i}_{C_i}$ to $\mathcal{L}^{P_j}_{C_j}$, denoted by $\mathcal{L}^{P_i}_{C_j} \xrightarrow{A^1_{i,j}} \mathcal{L}^{P_j}_{C_j}$, if the following conditions hold:
	
	\begin{enumerate}
	\item $\mathcal{L}^{P_i}_{C_i}$ and $\mathcal{L}^{P_k}_{C_k}$ are transformationally isomorphic, i.e. $\mathcal{L}^{P_i}_{C_i} \overset{0}{\sim} \mathcal{L}^{P_k}_{C_k}$, through some arrow $A^0_{i,k}:\langle T_{i,k} \rangle$
	\item The set of all valid input state assignments for $\mathcal{L}^{P_k}_{C_k}$ is a subset of the range of $\mathcal{L}^{P_l}_{C_l}$, i.e. $T_{i,k}(\mathcal{S}_{i,\perp_i}) \subseteq Im(\mathcal{S}_{l,\perp_l})$.
	\end{enumerate}
	
	Here, $\mathcal{A}^1_{i,j}$ is the set of all arrows of type-$1$ that can exist between $\mathcal{L}^{P_i}_{C_i}$ and $\mathcal{L}^{P_j}_{C_j}$, one for every sub-program $\mathcal{L}^{P_k}_{C_k}$ of $\mathcal{L}^{P_j}_{C_j}$ such that $\mathcal{L}^{P_i}_{C_i} \overset{0}{\sim} \mathcal{L}^{P_k}_{C_k}$. The existence of $A^1_{i,j}$ establishes a \textbf{Sub-structure Transformational Isomorphism} between $\mathcal{L}^{P_i}_{C_i}$ and $\mathcal{L}^{P_j}_{C_j}$, denoted by $\mathcal{L}^{P_i}_{C_i} \overset{1}{\sim} \mathcal{L}^{P_j}_{C_j}$.\\
	\end{mydef}  
	
	\begin{figure}[htp]
	\centering
	\begin{tikzpicture}
	\node (A) {$\mathcal{L}^{P_i}_{C_i}:\mathcal{S}_{i,\perp_i} \to \mathcal{S}_{i,\perp_i}$};
	\node (B) [below of=A, node distance = 2cm] {$\mathcal{L}^{P_j}_{C_j}: \mathcal{S}_{j,\perp_j} \to \mathcal{S}_{j,\perp_j}$};
	
	\draw[->] (A) to node {$A^1_{i,j}$} (B);
	\end{tikzpicture}
	\caption{Diagrammatic representation of type-$1$ arrow} 
	\label{fig:type1arrowDef}
	\end{figure}
	
	Similar to a type-$0$ arrow, a more detailed diagram for type-$1$ arrow is given in Fig. \ref{fig:type1arrowDef}. The two conditions mentioned in Def. \ref{def:type1arrow} ensure that for no execution of $\mathcal{L}^{P_l}_{C_l}$ on any valid input, do we lose some valid input for $\mathcal{L}^{P_k}_{C_k}$ and hamper our transformational isomorphism. In other words, if $\mathcal{L}^{P_l}_{C_l}$ was such that it mapped even a single element in the domain of $\mathcal{L}^{P_k}_{C_k}$ to $\{\perp\}$, then the latter would no longer be able to operate on it (or more specifically, $\mathcal{L}^{P_k}_{C_k}$ would not be able to get the machine out of this undefined state) and the isomorphism between $\mathcal{L}^{P_i}_{C_i}$ and $\mathcal{L}^{P_k}_{C_k}$ would cease to exist. \\
	
	At this stage, it becomes important to study some fundamental properties of type-$1$ arrows before we move on to complex ones. Assume that all arrows mentioned below exist, unless some contraints are specified otherwise.
	
	\begin{enumerate}
	\item There is no such thing as an inverse arrow for $A^1_{i,j}$, in general, for the programs $\mathcal{L}^{P_i}_{C_i}$ and $\mathcal{L}^{P_j}_{C_j}$. If such an arrow existed, it would actually have proved that $\mathcal{L}^{P_i}_{C_i}$ is isomorphic to some sub-program of $\mathcal{L}^{P_j}_{C_j}$ and viceversa, which would imply that $\mathcal{L}^{P_i}_{C_i} \overset{0}{\sim} \mathcal{L}^{P_j}_{C_j}$.
	  
	\item The existence of $A^0_{i,j}$ implies the existence of $A^1_{i,j}$ between $\mathcal{L}^{P_i}_{C_i}$ and $\mathcal{L}^{P_j}_{C_j}$, because we can always write $\mathcal{L}^{P_j}_{C_j} = (id_\emph{True} \circ (\mathcal{L}^{P_j}_{C_j} \circ id_\emph{True}))$. The vice versa is not true, for reasons similar to point 1.
	
	\item If we have $\mathcal{L}^{P_i}_{C_i} \overset{0}{\sim} \mathcal{L}^{P_j}_{C_j}$ and $\mathcal{L}^{P_j}_{C_j} \overset{1}{\sim} \mathcal{L}^{P_k}_{C_k}$, then we must have $\mathcal{L}^{P_i}_{C_i} \overset{1}{\sim} \mathcal{L}^{P_k}_{C_k}$. Similarly, if $\mathcal{L}^{P_i}_{C_i} \overset{1}{\sim} \mathcal{L}^{P_j}_{C_j}$ and $\mathcal{L}^{P_j}_{C_j} \overset{0}{\sim} \mathcal{L}^{P_k}_{C_k}$, then $\mathcal{L}^{P_i}_{C_i} \overset{1}{\sim} \mathcal{L}^{P_k}_{C_k}$ must hold as well. We call this property a \emph{Cross Transitivity} between type-$0$ and type-$1$ arrows.
	
	\item If we denote the set of all valid input state assigments for $\mathcal{L}^{P_i}_{C_i}$ by $[\mathcal{S}_i]_C$, and if we can write $\mathcal{L}^{P_j}_{C_j} = (id_{C_1} \circ (\mathcal{L}^{P_i}_{C_i} \circ id_{C_2}))$ such that $C \vdash C_2$ and $C_1 \vdash C$, then $\mathcal{L}^{P_i}_{C_i} \overset{1}{\sim} \mathcal{L}^{P_j}_{C_j}$.  
	
	\item The relation $\overset{1}{\sim}$ is reflexive, i.e. $\mathcal{L}^{P_i}_{C_i} \overset{1}{\sim} \mathcal{L}^{P_i}_{C_i}$ for all $\mathcal{L}^{P_i}_{C_i} \in \mathcal{P}$. This follows directly from the reflexivity of type-$0$ arrows.
	
	\item The relation $\overset{1}{\sim}$ is transitive, i.e. if $\mathcal{L}^{P_i}_{C_i} \overset{1}{\sim} \mathcal{L}^{P_j}_{C_j}$ and $\mathcal{L}^{P_j}_{C_j} \overset{1}{\sim} \mathcal{L}^{P_k}_{C_k}$, then we must have $\mathcal{L}^{P_i}_{C_i} \overset{1}{\sim} \mathcal{L}^{P_k}_{C_k}$. This also follows directly from the transitivity of type-$0$ arrows as well as the cross-transitivity of type-$0$ and type-$1$ arrows. Thus, all diagrams containing arrows of type-$1$ must necessarily commute.
	
	\item The above two points, along with the non-existence of an inverse arrow, proves that $\overset{1}{\sim}$ induces a partial order on the elements of $\mathcal{P}$. This ordered collection, denoted by $(\mathcal{P},\overset{1}{\sim})$, compares all those pairs of programs, one of which is sub-structure isomorphic to the other.  
	\end{enumerate}  
	
\begin{figure}[htp]
	\centering
	\begin{tikzpicture}
	\draw[dashed] (-4,0) circle(2.5);
	\draw[dashed] (4,0) circle(2.5);
	\draw[dashed] (0,4) circle(2.5);
	
	\node at (0,6) (P1) {$\mathcal{L}^{P_1}_{C_1}$};
	\node at (2,4) (P2) {$\mathcal{L}^{P_2}_{C_2}$};
	\node at (0,2) (P3) {$\mathcal{L}^{P_3}_{C_3}$};
	\node at (-2,4) (P4) {$\mathcal{L}^{P_4}_{C_4}$};
	
	\node at (4,2) (P5) {$\mathcal{L}^{P_5}_{C_5}$};
	\node at (2,0) (P6) {$\mathcal{L}^{P_6}_{C_6}$};
	\node at (4.3,0) (P7) {$\mathcal{L}^{P_7}_{C_7}$};
	\node at (4,-2) (P8) {$\mathcal{L}^{P_8}_{C_8}$};
	\node at (6,0) (P9) {$\mathcal{L}^{P_9}_{C_9}$};
	
	\node at (-4,-2) (P10) {$\mathcal{L}^{P_{10}}_{C_{10}}$};
	\node at (-4.8,1.5) (P11) {$\mathcal{L}^{P_{11}}_{C_{11}}$};
	\node at (-2.2,0.5) (P12) {$\mathcal{L}^{P_{12}}_{C_{12}}$};
	
	\draw[->] (P1) to node {$A^{0}_{1,2}$} (P2);
	\draw[->] (P1) to node {$A^{0}_{1,4}$} (P4);
	\draw[->] (P2) to node {$A^{0}_{2,3}$} (P3);
	
	\draw[->, bend left] (P2) to node {$A^{1}_{2,5}$} (P5);
	\draw[->, bend right] (P3) to node[below] {$A^{1}_{3,6}$} (P6);
	
	\draw[->] (P5) to node {$A^{0}_{5,6}$} (P6);
	\draw[->] (P7) to node {$A^{0}_{7,6}$} (P6);
	\draw[->] (P8) to node[right] {$A^{0}_{8,7}$} (P7);
	\draw[->] (P7) to node {$A^{0}_{7,9}$} (P9);
	
	\draw[->, bend left] (P7) to node {$A^{1}_{7,10}$} (P10);
	
	\draw[->] (P10) to node {$A^{0}_{10,11}$} (P11);
	\draw[->] (P12) to node {$A^{0}_{12,10}$} (P10);
	
	\draw[->, bend right] (P4) to node[above] {$A^{1}_{4,11}$} (P11);
	\end{tikzpicture}
	\caption{Reduction of type-$0$ arrows in a commuting diagram}
	\label{fig:type01diag}
	\end{figure}
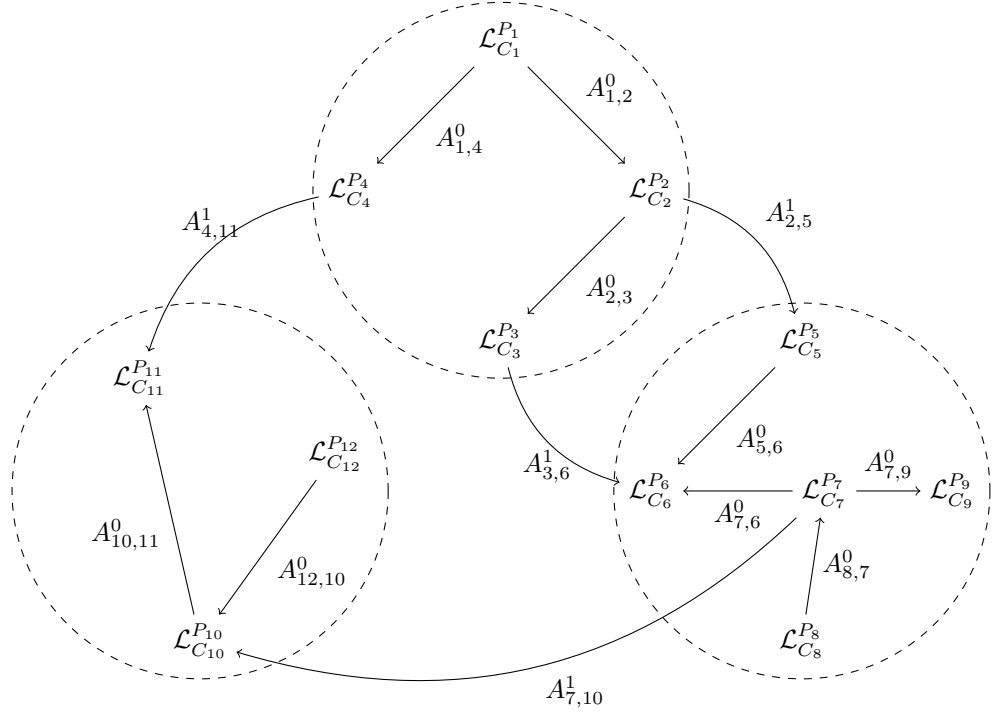

	Having studied these properties, we note that every diagram that commutes, containing arrows of type-$0$ between programs, can be converted into a diagram consisting of type-$1$ arrows only, just by relabelling each $A^0_{i,j}$ to $A^1_{i,j}$. Also, if a diagram contains both kinds of arrows, it can be reduced to a diagram containing only type-$1$ arrows (using cross transitivity) by clustering all isomorphic programs together and adding a type-$1$ arrow between two clusters if some program in the first cluster is sub-structure transformationally isomorphic to some program in the second cluster. This way, addition of type-$0$ arrows to a diagram containing type-$1$ arrows adds no extra information upto transformational isomorphism. The diagram in Fig. \ref{fig:type01diag} illustrates this reduction by clustering isomorphic programs inside dashed circles.\\
	
	The representation of $\mathcal{L}^{P_j}_{C_j}$ as a composition of three functions, or equivalently, concatenation of three programs, may not be unique. All we have to do is find some representation in which the required sub-structure isomorphism can be established. Let us look at a particularly interesting case, where $\mathcal{L}^{P_i}_{C_i}$ does not halt for some input. This means that $\mathcal{L}^{P_j}_{C_j}$ contains a fragment of code, isomorphic to $\mathcal{L}^{P_i}_{C_i}$, which exhibits non-terminating behavior for the corresponding input, obtained by the appropriate transformation function. The second condition in Def. \ref{def:type1arrow} assures that such an input is infact a valid input for $\mathcal{L}^{P_j}_{C_j}$ and hence, we are sure that $\mathcal{L}^{P_j}_{C_j}$ can enter indefinite computation. More formally, if any non-terminating program is sub-structure isomorophic to a given program $\mathcal{L}^{P_j}_{C_j}$, then this program must also be non-terminating for some set of inputs. The undecidability of determining the truth preservation order and limit of a function renders the problem of finding if such a program exists for $\mathcal{L}^{P_j}_{C_j}$ undecidable as well.\\
	
	The existence of type-$1$ arrows between programs, when visualized in the form of a directed acyclic (because of antisymmetry) graph, similar to the diagrams seen previously, provides a mechansim to enumerate all programs that are type-$1$ isomorphic to some given program. If we treat programs as nodes in this graph and type-$1$ arrows as directed edges between these nodes, then a breadth-first traversal, started at any node, say $\mathcal{L}^{P_1}_{C_1}$, would enumerate all programs that are type-$1$ isomorphic to $\mathcal{L}^{P_1}_{C_1}$. Seen as a whole, if we denote this graph by $\mathbb{G}_1$, then the connected components in this graph constitute the sets of mutually type-$1$ isomorphic programs. The non-existence of any type-$1$ arrow between programs across these components renders programs in one component \emph{non type-$1$ reachable} from the ones in the other component. A special component that is formed here is the one containing infinite loops. We denote this cluster by $\mathbb{H}_1$. Any program, $\mathcal{L}^{P}_{C}$ belonging to $\mathbb{H}_1$, must be sub-structure transformationally isomorphic to an infinite loop, $\mathcal{L}^{id}_{\emph{True}}$, which means that it must contain some non-terminating component. Even if $\mathcal{L}^{P}_{C}$ has one valid input state assignment for which indefinite computation is inevitable, it is sufficient for this program to fall in this component. Thus, while all programs in $\mathbb{H}_1$ necessarily show non-terminating behavior on atleast one input provided to them. However, this does not mean that $\mathbb{G}_1 \backslash \mathbb{H}_1$ contains only those programs that halt in finite time. The only programs in $\mathbb{G}_1 \backslash \mathbb{H}_1$ that do not halt are the ones that enter the undefined state some time during their execution and then cease to come out of this state, rendering the machine perform bogus computations forever. The program corresponding to $id_\perp$ must be type-$1$ isomorphic to all such \emph{outliers} of $\mathbb{G}_1 \backslash \mathbb{H}_1$, and hence, it forms yet another connected component in $\mathbb{G}_1$, say $\mathbb{K}_1$. Hence, the programs, now in $((\mathbb{G}_1 \backslash \mathbb{H}_1) \backslash \mathbb{K}_1)$ are guaranteed to halt in finite time. \\
	
	Every computable functions has an equivalent Turing Machine, and hence, a corresponding computer program. Thus, if $\mathcal{L}^{P_i}_{C_i} \overset{1}{\sim} \mathcal{L}^{P_j}_{C_j}$, then the computability of $\mathcal{L}^{P_j}_{C_j}$ implies computability of $\mathcal{L}^{P_i}_{C_i}$. A far more fascinating consequence is that given the computability of $\mathcal{L}^{P_j}_{C_j}$, we can be sure that $\mathcal{L}^{P_j}_{C_j}$ is atleast as hard to compute as $\mathcal{L}^{P_i}_{C_i}$. This implies that $\mathcal{L}^{P_i}_{C_i}$ is mapping reducible to $\mathcal{L}^{P_j}_{C_j}$, i.e. the complexity class of $\mathcal{L}^{P_i}_{C_i}$ is a subclass of the complexity class of $\mathcal{L}^{P_j}_{C_j}$. For example, if $\mathcal{L}^{P_j}_{C_j}$ is in NP, then $\mathcal{L}^{P_i}_{C_i}$ is in NP, but not the viceversa. More precisely, it is the complexity class of the transformation function, $T_{i,k}$ that decides the complexity class of $\mathcal{L}^{P_j}_{C_j}$. If $T_{i,j}$ is computable in polynomial time, then $\mathcal{L}^{P_i}_{C_i}$ is polynonial time reducible to $\mathcal{L}^{P_j}_{C_j}$. Thus, Turing machine reducibility, in general, is a direct consequence of the existence of a type-$1$ arrows. However, the inverse may not be true. We will capture of notion of generalized Turing machine reducibility between programs through the existence of type-$2$ arrows in the next subsection. 
	 
	\subsection{Type-$2$ arrow : Turing machine reducability}
	
	To some extent, both type-$0$ and type-$1$ arrows capture a weak equivalence, since the true semantic equivalence is very hard to achieve. For example, the problem of finding independent set of vertices in a graph is polynomial time reducible to the problem of determining if a given Boolean formula has a satisfying assignment of its literals. But this does not mean that a type-$1$ arrow exists between the two, because the kind of control structures used in both the algorithms is entirely different. It is the result of one that we use to derive the result for the other, but we are really not finding an independent set inside the Boolean formula while checking satisfying assignments per se. At this point, one may say that the program to solve Boolean satisfiability can be written as $(\mathcal{L}^{P_3}_{C_3} \circ (\mathcal{L}^{P_2}_{C_2} \circ \mathcal{L}^{P_1}_{C_1}))$, where $\mathcal{L}^{P_1}_{C_1}$ first converts the given instance of SAT into an instance of independent set, then $\mathcal{L}^{P_2}_{C_2}$ solves independent set problem on that instance, and finally, $\mathcal{L}^{P_3}_{C_3}$ converts the output so obtained into the form required by SAT. However, this only proves that $\mathcal{L}^{P_2}_{C_2}$ is transformationally isomorphic to the program to solve independent set problem on a graph structurally isomorphic to that obtained as the output of $\mathcal{L}^{P_1}_{C_1}$. We still do not have a sub-structure isomorphism between SAT and independent set, and hence, polynomial time reducibility, or Turing machine reducibility, in general, does not imply the existence of type-$1$ arrows. Thus, the notion of Turing machine reducibility must be taken into account, because it is precisely this relation which encapsulates how one program is related to other. \\
	
	\begin{mydef}
	\label{def:type2arrow}
	Let $P_i, P_j \in \mathcal{P}$ be two computer programs, and, $TM_i$ and  $TM_j$ be two Turing machines that accept languages $L(P_i)$ and $L(P_j)$, respectively. Then there exists an \textbf{arrow of type-$2$}, $A^2_{i,j} \in \mathcal{A}^2_{i,j}$ between $P_i$ and $P_j$, denoted by $P_i \xrightarrow{A^2_{i,j}} P_j$, if the language $L(P_i)$ is Turing reducible to language $L(P_j)$, i.e. $L(P_i)$ is decidable relative to $L(P_j)$.\\ 
	
	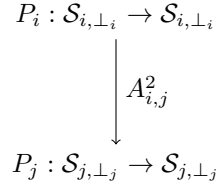
\begin{figure}[htp]
	\centering
	\begin{tikzpicture}
	\node (A) {$P_i:\mathcal{S}_{i,\perp_i} \to \mathcal{S}_{i,\perp_i}$};
	\node (B) [below of=A, node distance = 2cm] {$P_j: \mathcal{S}_{j,\perp_j} \to \mathcal{S}_{j,\perp_j}$};
	
	\draw[->] (A) to node {$A^2_{i,j}$} (B);
	\end{tikzpicture}
	\caption{Diagrammatic representation of type-$2$ arrow} 
	\label{fig:type2arrowDef}
	\end{figure}
	
	Here, $\mathcal{A}^2_{i,j}$ is the set of all arrows of type-$2$ that can exist between $P_i$ and $P_j$, one for every pair of Turing machines that accept the languages $L(P_i)$ and $L(P_j)$ at least, respectively. The existence of $A^2_{i,j}$ establishes a \textbf{Turing machine reducibile isomorphism} between $P_i$ and $P_j$, denoted by $P_i \overset{2}{\sim} P_j$.\\ 
	\end{mydef}
	
	Note that the definition asserts the Turing machines corresponding to $P_i$ and $P_j$ to accept at least $L(P_i)$ and $L(P_j)$, respectively. This means that the machines, $TM_i$ and $TM_j$, respectively, accept languages such that $L(P_i) \subseteq L(TM_i)$ and $L(P_j) \subseteq L(TM_j)$. However, we do not require $L(TM_i)$ to be Turing reducible to $L(TM_j)$ for a type-$2$ arrow to exist. This way, we are really not talking of $TM_i$ and $TM_j$ to strictly correspond to $P_i$ and $P_j$, respectively, but in some sense, their execution simulates the behavior of these programs, in disguise. More precisely, the true nature of $P_i$ and $P_j$ is not necessarily simulated in the exact sense these programs are coded, but possibly in some other way such that the transformations these programs provide to their input state assignments is replicated by the Turing machines on the same set of inputs.\\
	
	Turing-reducibility talks of decidability relative to a language, which essentially means if language $A$ is Turing-reducible to language, $B$, then given a Turing machine to solve $B$, we can always solve $A$. The condition of existence of a Turing machine to solve $B$ is hypothesized, for situations in which such a machine would, at least theoretically, not exist, using an oracle for $B$, which can be visualized as an external device that is capable of reporting whether any string $\omega$ is a member of $B$. If any Turing machine, $M$, has an additional capability to query oracles of this form, we call it an \emph{Oracle Turing machine}, denoted by $M^B$. The relative decidability of $A$ can now be explained by checking if $M^B$ decides $A$. This way, mapping reducibility becomes a special case of Turing reducibility and hence, type-$1$ arrows form a special case of type-$2$ arrows. The properties of type-$2$ arrows are, hence, similar to those of type-$1$ arrows, with some obvious modifications.\\
	
	Having studied the various kinds of arrows, we now give a formal method to represent computer programs as partial functions. Once we have our representation ready, a category of sequential programs can be formed using the arrows just described. Although we shown that every type-$0$ arrow can be converted into a type-$1$ arrow, and subsequently, into a type-$2$ arrow, we do not aim to form diagrams containing only arrows of type-$2$. The seemingly weak similarity, as captured by type-$0$ and type-$1$ arrows is sufficient strong enough to compare two programs based on the number of computations they perform over a given set of input state assignments. As was explained previously, important constructs, like indefinite looping, can be compared using these two arrows only, and we need to resort to something as strong as Turing reducibility. Hence, the diagrams may not be reduced to contain only type-$2$ arrows, unless specifically needed for some purpose.
	
	\section{Representing programs as partial functions}
		
		The isomorphism, as discussed previously, between computer programs is with respect to the truth preservation and order-preservation of the transformation function between their state variables. The encoding of any computable function in a suitable programming language makes it difficult to deduce the set of inputs for which the program continues its execution before halting. We also know that no such representation of these instructions exist, for if one did, then that representation would have provided a way to solve the so-called Halting problem. However, if we can somehow represent these instructions in a form so that at least specifying the set of valid state assignments becomes easy, then we have a model which imparts enough readability for us to continue with our categorization. The partial function approach fits best in this situation, and thus, we now learn a simple technique to represent any computer program using a partial function. The condition governing the selection of elements in the domain of partial functions on which the function is defined acts as a filter for selecting the valid state assignments for the corresponding computer program. Such a representation is unique upto membership in $EQ_{TM}$, i.e. the partial functions corresponding to the pair of Turing machines encoded in the language, $EQ_{TM}$, can be used interchangeably. \\
		
		Every sequential program has an advantage that it's execution can be visualized using a control flow graph. The basic elements of any control flow graph are rectangular blocks for individual statements of the program and diamond-shaped blocks for condition checking. The directed links from one block to another depict the data flow as well as the control flow in the program. For example, consider the code as given in Fig. \ref{Fig:condFunc2}. The individual blocks in the flowchart for this code can be replaced by equivalent partial functions and we compose these functions following the direction of links. Let $y \in A$ be the value used for initializing $x$. Then the partial function equivalent to the program in Fig. \ref{Fig:condFunc2} is given as
		\begin{equation}
		\bigg( id_{\emph{True}} \bigg( \mathcal{L}_{C_A}^{\mathcal{L}_{C_A \land \lnot C}^{f_2} \left( \mathcal{L}_{C_A \land C}^{f_1} \right)} \left( \beta_y \right) \bigg) \bigg) : \mathcal{S}_\perp \to \mathcal{S}_\perp
		\end{equation}
		
		Ignore the notations for the time being; they will be explained shortly. The code in Fig. \ref{Fig:condFunc} is equivalent to that in Fig. \ref{Fig:condFunc2}. The corresponding partial function for the former code is given as
		\begin{equation}
		\bigg( id_{\emph{True}} \bigg( \mathcal{L}_{C_A}^{\mathcal{T}_{C}^{f_1,f_2}} (\beta_y) \bigg) \bigg) : \mathcal{S}_\perp \to \mathcal{S}_\perp
		\end{equation}
		
		The two partial functions must also be equivalent to account for the equivalence in the two corresponding codes. Thus, we can give the following equivalence property:
		\begin{equation}
		\bigg( \mathcal{L}_{C_A}^{\mathcal{T}_{C}^{f_1,f_2}} \bigg) \equiv \bigg( \mathcal{L}_{C_A}^{\mathcal{L}_{C_A \land \lnot C}^{f_2} \left( \mathcal{L}_{C_A \land C}^{f_1} \right)} \bigg)
		\end{equation}
		
		Most of what is being talked of right now will make more sense once we introduce the symbols $\mathcal{T}$ and $\beta$ in our discussion. These represent if-conditionals and assignments, respectively. A more formal discussion will follow shortly. We exploit the fact that only four kinds of special partial functions are enough for representing every sequential program, along with the standard representations for mathematical transformations. These four functions are : conditional/unconditional identity function, if-conditional, while-loop and assignments. We have already studied the notations for identity function and while-loop, and hence, we will not repeat that here. Let us see the notations for assignments and if-conditionals in more detail.
		
		\section{The assignment statement}
		The assignment statement assigns to some variable a value, which is assumed to be constant and immutable. We do not allow expressions in our assignment statement. Thus, $x \gets 2$ is a valid assignment statement but not $x \gets x+3$ or $x \gets 6-4$ (The latter is not allowed for simplicity in representation. An equivalent set of statements to mimic this behavior is $t1 \gets 6$, $t2 \gets 4$, $x \gets t1-t2$, of which only the first two are valid assignment statements. The third one is a mathematical function). We only allow the statement $x \gets y$ to represent a valid assignment statement if $x$ belongs to the set of $L$-values and $y$ belongs to the set of $R$-values (and is not an expression) for the given model of computation. This assignment is represented, in shorthand, by $\beta_y(x)$. Thus, $\beta_y(x) = id(y) = y$ if and only if $x$ is an $L$-value in our model and $y$ is an $R$-value, and not an expression. \\
		
		We extend the definition of $\beta_y(x)$ to account for mutiple assignments at the same time (while preserving the sequential nature of execution) as well as treat functions as first-class members, so that one function can be assigned to other (similar to beta renaming in lambda calculus). We treat the subscript as well the argument to $\beta$ as ordered tuples or arbitrary arity, which will be evident from the situation. The tuples follow the basic recursive property that $(x,y)=(x):(y)=(x,y):()=():(x,y)$, where $()$ is the empty tuple and $:$ represents concatenation. This way, we now define our $\beta$ functional as: 
		
		\begin{equation}
		\beta_{(y_1,y_2,y_3,\dotsc,y_k)}(x_1,x_2,x_3,\dotsc,x_k) = \left\lbrace
		\begin{split}
		x_1 \gets y_1 \\
		x_2 \gets y_2 \\
		x_3 \gets y_3 \\
		\vdots \\
		x_k \gets y_k
		\end{split}
		\right\rbrace
		\end{equation}
		
		The statements in the big brace above are executed from top to bottom, and hence, this definition does not violate the conditional of sequential execution. A recursive definition can also be given as
		
		\begin{equation}
		\beta_{(y_1,y_2,y_3,\dotsc,y_k)}(x_1,x_2,x_3,\dotsc,x_k) = \left\lbrace
		\begin{split}
		x_1 \gets y_1 \\
		\beta_{(y_2,y_3,\dotsc,y_k)}(x_2,x_3,\dotsc,x_k)
		\end{split}
		\right\rbrace
		\end{equation}
		which can also be written as
		
		\begin{equation}
		\beta_{(y_1,y_2,y_3,\dotsc,y_k)}(x_1,x_2,x_3,\dotsc,x_k) = (y_1):\beta_{(y_2,y_3,\dotsc,y_k)}(x_2,x_3,\dotsc,x_k)
		\end{equation}
		
		As an example, let $x,y,z$ be three variables in our program, which have been initialized to values $2,3,4$ respectively. Then the functional $\beta_{(5,2,0)}(x,y,z)$ transforms these variables to assume their new values as $5,2,0$ respectively. In principle, $\beta$ is actually a functional belonging to an infinite family of functionals defined below:
		
		\begin{equation}
		\beta_{(y_1,\dotsc,y_k)}(x_1,\dotsc,x_k) \in \underbrace{\left[ \overbrace{\mathcal{S}_\perp \times \dots \times \mathcal{S}_\perp}_{\text{$n$-times}} \to \overbrace{\mathcal{S}_\perp \times \dots \times \mathcal{S}_\perp}_{\text{$n$-times}} \right]}_{\text{Set of all functions from $\mathcal{S}_\perp^n$ to itself}}  \quad \text{ for } 0< k < n < \infty 
		\end{equation}
		
		This way, since $\beta$ acts as a partial function. Some notational abuse can be introduced at this stage, for the sake of brevity. In the subscript of $\beta$, if the number of $R$-values is less than the number of $L$-values in the argument, then assignment the last $R$-value to the remaining $L$-values. For example, $\beta_{(1)}(x,y,z)$ can stand for assigning $1$ to $x,y,z$ each, instead of writing $\beta_{(1,1,1)}(x,y,z)$ for the same. Similarly, we can have
		
		\begin{equation*}
		\beta_{(1,2)}(x,y,z) = \left\lbrace
		\begin{split}
		x \gets 1 \\
		y \gets 2 \\
		z \gets 2
		\end{split}
		\right\rbrace
		\end{equation*} 
		
		Mutliple occurances of the same variable in the argument to $\beta$ can be reduced, without any loss of meaning, as follows:
		
		\begin{equation*}
		\beta_{(1,2,3)}(x,y,x) = \left\lbrace
		\begin{split}
		x \gets 1 \\
		y \gets 2 \\
		x \gets 3
		\end{split}
		\right\rbrace =
		\left\lbrace
		\begin{split}
		y \gets 2 \\
		x \gets 3
		\end{split}
		\right\rbrace = \beta_{(2,3)}(y,x) 
		\end{equation*} 
		
		The order of elements in the tuples does not matter as long as the relative order of the $L$-values and $R$-values does not change. Thus, $\beta_{(2,3)}(y,x) = \beta_{(3,2)}(x,y)$. In case any variable is assigned an undefined value, which may happen if the assignment is not valid due to type inconsistencies in typed languages, then the machine enters the undefined state immediately and all subsequent assignments make no difference to the state of the machine. Hence,
		
		\begin{equation}
		\beta_{(y_1,\dotsc,y_{i-1},\perp,y_i,\dotsc,y_k)} \equiv id_\perp
		\end{equation} 
		
		We can also incorporate the concept of variable renaming, similar to that in lambda calculus, through this $\beta$ notation. If $f_1,f_2,\dotsc,f_k$ are lambda expressions, in which $x_1,x_2,\dotsc,x_k$, respectively are bound variables and we choose $y_1,y_2,\dotsc,y_k$ such that none of the $y_i$'s belongs to the set of bound variables for $f_i$, then the following notation can compactly represent renaming operations on these $k$ lambda expressions. Assume that $f[y/x]$ stands for the renaming all occurances of the bound variable $x$ to $y$ in the function $f$.
		
		\begin{equation}
		\beta_{([y_1/x_1],[y_2/x_2],\dotsc,[y_k/x_k])}(f_1,f_2,\dotsc,f_k) = \left\lbrace
		\begin{split}
		f_1[y_1/x_1] \\
		f_2[y_2/x_2] \\
		\vdots \\
		f_k[y_k/x_k]
		\end{split}
		\right\rbrace
		\end{equation}
		
		So much for the assignment statements, the next important notation we study is the one for if-conditionals. 
		
		\subsection{The if-conditional}
		
		The if-conditional is perhaps the most important and frequently encountered statement in any computer program. It forms the core behind any computation which is performed only in a selected few situations, this selection being done through the condition specified. The most general form of an if-statement is given as : \emph{if $C$ is True, then set $x \gets f_1(x)$, else set $x \gets f_2(x)$}. A shorthand for this is given by $\mathcal{T}_C^{f_1,f_2}$. We use the symbol $\mathcal{T}$ because it resembles the shape of the control flow graph when an if-condition is executed. Note the order of functions in the superscript. The first one is executed when the condition in the subscript is \emph{True}, while the second one executes otherwise. It is worth noting that $\mathcal{T}_C^{f_1,f_2}$ has only a single entry point and a single exit point, which allows us to treat it as a single transformation of the state variables. Furthermore, it is defined for all elements in $\mathcal{S}_\perp$, and hence, is a total function. An obvious modification can be done with $\mathcal{T}_C^{f_1,f_2}$ by negating the condition, that is
		
		\begin{equation}
		\mathcal{T}_C^{f_1,f_2} \equiv \mathcal{T}_{\lnot C}^{f_2,f_1}
		\end{equation}  
		
		The order of functions in the superscript reverses in this case. An interesting way to represent while-loops using this conditional can be done as following.
		
		\begin{equation}
		\begin{split}
		L(x) &=\mathcal{T}_C^{(L \circ f_1)(x),id_{\emph{True}}}\\
		& \equiv \mathcal{L}_C^{f_1}(x)
		\end{split}
		\end{equation}
		
		In other words, if we express this in terms of a lambda expression as given below, then $\mathcal{L}_C^{f_1}(x)$ forms the least fixed point for the recursive function $Y(G)$, where $Y$ represents the $Y$-combinator.  
		
		\begin{equation}
		L(x) \equiv \underbrace{\lambda f .\lambda x. \text{ (if $C(x)$ is True, then $(f(f_1(x)))$ else $x$} )}_{\text{Lets us call it $G$} }(L)(x)
		\end{equation}
		
		This way, a while-loop becomes a recursive counterpart of the if-conditional. The while-loop can also be used to represent conditional identity functions by observing that $\mathcal{L}_C^{id} \equiv id_{\lnot C}$. On similar lines, the if-conditional can be used for the same purpose by observing that $id_C \equiv \mathcal{T}_C^{id,id_\perp}$. Thus, we can actually represent all programs using only three functions: $\beta,\mathcal{L}$ and $\mathcal{T}$. However, for the sake of brevity, we will use $id$ in our discussions as well.\\
		
		With all the necessary tools available, the arrows to capture similarity between programs will be put to group similar programs together. The representation using partial functions allows us to extract a Boolean condition $C \in \mathcal{S_\perp}$ such the given program $P$ can be written as $P : [\mathcal{S}_\perp]_C \to \mathcal{S}_\perp$ in a way that $C$ filters all those state assignments in $S_\perp$ that cause $P$ to halt. Although finding such a $C$ is an undecidable task, it is convinient to assume its extraction for mathematical writing and conceptualization.
	
	\section{A category of sequential algorithms}
	
	This section uses the theory developed so far to present a possible categorization of sequential programs, with the help of their representations using partial functions. Before moving forward with this, we must emphasize that the categorization discussed here is one of the many possible ways to form a category of computer programs. The basis of our categorization is essentially the existence of a truth preserving mapping from one program to another, which generalizes easily to sub-structure isomorphism and Turing machine reducibility. Two examples of categorizations, similar to the one we will shortly develop, are given as follows.\\
	
	Let $P$ be a computer program, which contains basic instructions, like assignments, mathematical operations, logical operations and jump statements (conditional and unconditional) only (user input and output statements can also appear). We can always represent this program using a flowchart, comprising of blocks for each instruction and arrows from one block to another, depicting the control flow. These arrows can also go from a set of blocks to itself, in case some part of the program requires looping. If we have an arrow connecting block-$1$ to block-$2$, and another connecting block-$2$ to block-$3$, then this means that the control flows from block-$1$ to block$2$ and then to block-$3$, implying that eventually, the control will flow from block-$1$ to block-$3$ as well, through subsequent tranformations. This hints towards a control-flow reachability using the transitivity of these arrows, and hence, composition of two arrows, obtained by concatenating the blocks connected by them in such a way that they get executed in the order of concatenation, can be used to establish this relation. If we now view our instructions inside one block to be representing an object, then all admissible combinations of instructions inside one block constitute various objects that can constitute a part of any given program. By admissible combinations, we mean any combination (rather permuation in most cases) of instructions which can form a basic block, i.e. a block in which the jump instruction, if present, is the last instruction, so that if at any point the program execution reaches this block, all the instructions will be necessarily executed.  The arrows connecting these blocks, also connect the corresponding objects. These arrows, being reflexive and transitive, allow for the existence of identity arrows and composite arrows between objects. If we represent the set of all objects, defined as above, as $\mathcal{O}$ and the set of all arrows, connecting different blocks in a program, as $\mathcal{A}$, then $(\mathcal{O},\mathcal{A})$ forms a category. An interesting fact about this formulation is that every diagram represents a complete computer program, for which the topological alignment of objects and arrows provides the corresponding control-flow. Thus, all the sub-diagrams of a given diagram correspond to programs that are type-$1$ isomorphic to the program for the original diagram. However, if two diagrams are structurally isomorphic, then there must exist a bijection between the sets of objects as well as the sets of arrows in the two diagrams. The existence of this computable bijection establishes a type-$0$ isomorphism between the corresponding programs. Although this categorization looks appealing, it fails to capture type-$2$ isomorphism, and in most cases, type-$0$ isomorphism as well.\\ 
	
	Yet another categorization can be obtaining by visualizing state assignments as objects and computer programs as arrows. In such a model, we say that there exists an arrow between two objects if there exists a computable function which will transform the state assignment of object-$1$ into the state assignment of object-$2$. Quite intuitively, an identity arrow is the simple identity function here, and the composition of arrows is the concatenation of the corresponding programs. Since every state assignment is a countable collection of bits, each assuming a value $0$ or $1$, an arrow between two such assignments simply flips on of the bits that do not match. If we represent this category as $C$, then a more interesting category that arises here is the product category, 
	
	\begin{equation}
	C^k = \underbrace{C \times C \times \dots \times C}_{k-\text{times}}
	\end{equation}
	
	The objects in this product category are collections of $k$-state assignments, which means that the arrows now represent computable bijective functions, mapping each of these $k$-assignments to its corresponding assignment in the other object. The existence of this arrow is evidence to the existence of such a computable bijection, whose domain is the set of state assignments in the first object and whose range is the set of state assignments in the second object. The different arrows that exist between the same pair of objects represent programs that are type-$0$ reducible to each other. Also, since every object contains exactly $k$-state assignments, there always exists a bijection between these two objects and hence, any two objects are reachable to each other through the existence of arrows which represent programs that are type-$0$ isomorphic to the identity arrow for this product category. This transformational isomorphism is also present between arrows in $C^k$ and those in $C^{k+n}$, for all $n > 0$, as is apparent from Def. \ref{def:type0arrow}. \\
	
	Having looked at these two examples, we are convinced that there is no unique way to form a category when dealing with computer programs. The type of categorization, as can be obtained using the arrows discussed in previous sections, focusses on representing programs as objects and the arrows as the relations between these objects, so that the isomorphism, or reducibility, can be directly inferred from the commutative diagrams. The computer programs are all assumed to have the same domain and range, denoted by $\mathcal{S}_\perp$. The arrows between programs will be either type-$0$, type-$1$ or type-$2$, depending on the situation. We saw that these arrows can be composed together using the laws of transitivity and cross-transitivity, and that this composition is associative, i.e. $A_1 \circ (A_2 \circ A_3) = (A_1 \circ A_2) \circ A_3$, where $A_1, A_2$ and $A_3$ are arrows of the appropriate type. The existence of identity arrows between programs was also seen. We can assign to every arrow a source object and a target object, to be the partial functions corresponding to the two programs related by this arrow and hence, a formal \emph{typing} can be imparted in this context. This way, all the basic axioms of a category are satisfied and our partial functions along with the arrows form a category of sequential programs. \\
	
	An important result of such a categorization is the existence and encapturing of the sub-structure isomorphism between programs. For most applications, it may not be important to know the nature of actual calculation being performed. The analysis of computational complexity is such an application. Counting the number of steps as a function of input size is often independent of the actual computation performed which is why we talk not of an algorithm with a given complexity, but the set of all such algorithms. Type-$0$ arrows capture sub-structure isomorphism along with computational complexity, which is why type-$2$ arrows, which are purely based on Turing-machine reducibility, are directly implied from them. This categorization will be discussed in much greater detail in the final report, to follow. The next report will also contain interesting sub-categories, specially that of deterministic programs and terminating programs. The aim will be to use transfinite induction to study equivalence classes formed of programs. Another extension will be to study probabilistic algorithms and fit them into the category. Non-deterministic programs will also be explored and the categorization extended accordingly.

	\label{Bibliography}   
	\nocite{*}
	\bibliographystyle{acm}
	\bibliography{ref}
	
	\end{document}